\documentclass[12pt]{article}
\usepackage{amsfonts,amsmath,amssymb}
\usepackage{scalerel}[2016/12/29]
\usepackage{graphicx}
\usepackage{tikz}

\def\be{\begin{equation}}
\def\ee{\end{equation}}
\def\bea{\begin{eqnarray}}
\def\eea{\end{eqnarray}}

\def\sh{\hskip -0.05cm} 
\def\sb{\hskip -0.1cm} 
\def\sbb{\hskip -0.2cm}

\def\rme{\mathrm{e}}
\def\rmi{\mathrm{i}}
\def\scq{{\textsc q}}

\usepackage{amsmath}
\usepackage{epsfig,graphicx}
\usepackage{graphicx}
\usepackage{tikz}

\def\be{\begin{equation}}
\def\ee{\end{equation}}
\def\bea{\begin{eqnarray}}
\def\eea{\end{eqnarray}}

\def\rme{\mathrm{e}}
\def\rmi{\mathrm{i}}
\def\scq{{\textsc q}}

\begin{document}

\thispagestyle{plain}

\title{\bf\Large Lattice walk area combinatorics, some remarkable trigonometric sums and Ap\'ery-like numbers}
%\title{\Huge{Mapping the Calogero model on the Anyon model}}

\author{St\'ephane Ouvry$^*$ \  {\scaleobj{0.9}{\rm and}} Alexios P. Polychronakos$^\dagger$}

%\date{\today}

\maketitle

\begin{abstract}
\iffalse
We focus on remarkable trigonometric sums and their explicit rewriting in terms of binomial multiple  sums. They play a key role in the algebraic area enumeration of lattice random walks and are also important building blocks  of non trivial quantum models such as the Hofstadter  model. Explicit algebraic area enumeration formulae are proposed for various lattice walks in particular  the triangular lattice chiral walks introduced in \cite{poly}. An intriguing connection is also made with number theory and some classes of Ap\'ery-like numbers, the cousins of the Ap\'ery numbers which play a central role in irrationality  considerations for $\zeta(2)$ and $\zeta(3)$. 
\fi
Explicit algebraic area enumeration formulae are derived for various lattice walks generalizing  the canonical square lattice walk, and in particular for the triangular lattice chiral walk recently introduced by the authors. A key element in the enumeration is the derivation of some remarkable identities involving trigonometric sums --which are also important building blocks  of non trivial quantum models such as the Hofstadter  model-- and their explicit rewriting in terms of multiple binomial sums. An intriguing connection is also made with number theory and some classes of Ap\'ery-like numbers, the cousins of the Ap\'ery numbers which play a central role in irrationality considerations for $\zeta(2)$ and $\zeta(3)$.
\end{abstract}

\noindent
* LPTMS, CNRS,  Universit\'e Paris-Sud, Universit\'e Paris-Saclay,\\ {$\left. \right. $}\hskip 0.3cm 91405 Orsay Cedex, France; {\it stephane.ouvry@u-psud.fr}

\noindent
$\dagger$ Department of Physics, City College of New York, NY10031 and
the Graduate Center\\ {$\left. \right. $}\hskip 0.3cm of CUNY, New York, NY 10016, USA;
{\it apolychronakos@ccny.cuny.edu} 
\vskip 1cm

\vfill
\eject

\section{Introduction}

%{\color{red}I agree with all please do omit and modify. I made a few changes just above 3.2  and changed the numbering of the figures. }

The enumeration of random walks of given algebraic area on a two-dimensional lattice is  a hard and challenging  problem. The algebraic area is defined as the oriented area spanned by the walk as it traces the lattice. A unit lattice cell enclosed in the counterclockwise (positive) way has an area
$+1$, whereas when enclosed in the clockwise (negative) way it has an area $-1$. The total algebraic area is the area enclosed by the walk weighted by the winding number: if the walk winds around more than once, the area is counted with multiplicity.  
The combinatorics of such walks depend on the exact rule generating them and on the lattice geometry.
The canonical example is closed random walks on a square lattice. This problem can be mapped to the famous Hofstadter
model \cite{Hofstadter}  of a particle hopping on a square lattice pierced by a constant magnetic field, with the value of the magnetic field
playing a role analogous to the chemical potential for the area of the walk. Indeed, algebraic area enumerations are  mapped on quantum mechanical models since in quantum mechanics a magnetic field couples  to the area spanned by the  particle. 

An exact formula for the number of square lattice
walks of given length and algebraic area was only recently obtained in the form of nested binomial sums \cite{nous}.
The analysis  revealed some remarkable trigonometric sums to be key ingredients for the
 algebraic area  enumeration. 
They  are defined for $p$ and $q$ coprime positive integers as
\be{1\over q}\sum _{k=1}^q {b_{p/q}}^{{l_1}}(k){{{{b}}}_{p/q}}^{l_2}(k+1)\ldots {b_{p/q}}^{l_{j}}(k+j-1)\label{11}\ee
where ${b_{p/q}}(k)$ is a  trigonometric  function called spectral function which depends on the rational number $p/q$, and  $l_1,l_2,\ldots,l_j$ is a set of  positive or null integers. In the  algebraic area enumeration for square lattice walks   these integers are the parts in the compositions of the integer $n$, i.e., $n=l_1+\ldots+l_j$ and all $l_i$ positive,
with $n$ fixing the length of the walk. But we will consider more general lattice walks where some of the  $l_i$'s  can be null, in a way to be specified below.

In \cite{nous} the focus  was on the spectral function  
\be {{{{b}}}_{p/q}}(k)=\big(2\sin(\pi k p/q)\big)^2\label{22}\ee
 which encodes the Hofstadter  {dynamics}.
The  algebraic area enumeration was obtained  in part thanks to an explicit rewriting  of the trigonometric sum (\ref{11}), when evaluated for the Hofstadter spectral function (\ref{22}), in terms of the binomial multiple sums  
 \begin{align} 
 &{1\over q}\sum _{k=1}^q {{{{b}}}_{p/q}}^{{l_1}}(k){{{{b}}}_{p/q}}^{l_2}(k+1)\ldots {{{{b}}}_{p/q}}^{l_{j}}(k+j-1)=\sum_{A=-\infty\atop A\;{\it even}}^{\infty}\rme^{\rmi \pi A p/q} \label{33}\\&\sum_{k_3= -l_3}^{l_3 
 }\ldots \sum_{k_{j}= -l_j}^{l_{j}}{2l_1\choose l_1 +A/2+\sum_{i=3}^{j}(i-2)k_i}{2l_2\choose l_2  -A/2
 - \sum_{i=3}^{j}(i-1)k_i}\prod_{i=3}^{j}{2l_i\choose l_i+k_i}\nonumber
\end{align}
% where the exponential factor $\rme^{\rmi \pi A p/q}$ was  related to walks of algebraic area $A/2$ (see  section (\ref{myHof}) below).
Eq.~(\ref{33})  is  valid for any set of positive or null integers {$l_i$} with an $A$-summation range finite due to the first two binomials, where {$A$} appears.  In the specific case where the $l_i$'s are all positive  --as is the case for the square lattice walks algebraic area enumeration-- $A$ is restricted in the interval $[-2\lfloor (l_1+\ldots+l_j)^2/4\rfloor,\;2\lfloor (l_1+\ldots+l_j)^2/4\rfloor\;]$. When some of the $l_i$'s are null these bounds can be generalized  (see, e.g., the bounds in eq.~(\ref{even})). 

We  note that when we replace $\rme^{ \rmi \pi A p/q}$ by $1$ in (\ref{33}) we get the binomial identity
 \begin{align}
  &{2(l_1+\ldots+l_j)\choose l_1+\ldots+l_j}=\sum_{A=-\infty\atop A\;{\it even}}^{\infty}\label{333}\\&\sum_{k_3= -l_3}^{l_3 
 }\ldots \sum_{k_{j}= -l_j}^{l_{j}}{2l_1\choose l_1 +A/2+\sum_{i=3}^{j}(i-2)k_i}{2l_2\choose l_2 -A/2-\sum_{i=3}^{j}(i-1)k_i}\prod_{i=3}^{j}{2l_i\choose l_i+k_i}\nonumber
 \end{align}
 where the resulting binomial {in the LHS}\footnote{This binomial counting can be easily checked by first summing over $A$ and subsequently over the $k_i$'s, redefining them appropriately; see \cite{nous}.} will be interpreted later on as a factor contributing to the counting of lattice walks. Again, formula (\ref{333})  is  valid for any set of positive or null integers {$l_i$}; if the $l_i$'s are all positive the bounds on $A$ are as specified above.
  
We remark here that the trigonometric sum (\ref{11})  reduces to the binomial multiple sum  given in  (\ref{33})  in the case ${{{{b}}}_{p/q}}(k)=\big(2\sin(\pi k p/q)\big)^2$  only when $ l_1+\ldots+l_j<q$, i.e., for large enough  values of $q$. In view of the algebraic area enumeration of square lattice walks, where the algebraic area counting  ends up  being  the coefficient of $\rme^{\rmi \pi A p/q}$ {(see (\ref{b}) below)}, this constraint on $q$ {eliminates open} walks which could be confused with closed ones  by periodocity\footnote{Extrapolating (\ref{33}) as such  to  any value of $q\ge 1$ % --the algebraic area counting will end up  being (see below (\ref{b})) the coefficient of $\rme^{\rmi \pi A p/q}$, this regardless of any particular value of $p$, $q$ or of the $l_i$'s. 
would amount to  enforcing, for  any given integer $l$,  the identity  
$
\sum_{k=1}^{q} \rme^{2\rmi k \pi  p l / q} = 0
$
 even though this is valid only when $l$ is not a multiple of $q$
 (when $l$ is a multiple of $q$ the sum is actually equal to $q$).}.  %In what follows all results involving the trigonometric sums ${1\over q}\sum _{k=1}^q {{{{b}}}_{p/q}}^{{l_1}}(k){{{{b}}}_{p/q}}^{l_2}(k+1)\ldots {{{{b}}}_{p/q}}^{l_{j}}(k+j-1)$ will have to be understood in this way. 

In \cite{poly} we revisited the algebraic area enumeration of \cite{nous} and noted that it admits a statistical mechanical interpretation in terms of particles obeying generalized {\it exclusion statistics} \cite{Haldane} with exclusion parameter 
$g=2$ ($g=0$ for bosons, $g=1$ for fermions, higher $g$ means a stronger exclusion beyond Fermi). Other lattice
walks admit a similar interpretation with higher integer values of $g$.
We also introduced the notion of $g$-compositions where some zeros can be inserted at will inside the set of the $l_i$'s with the restriction that no more than $g-2$ zeros lay in succession.  The integer $n$ {admits} $g^{n-1}$ such compositions. In particular, $g\hskip -0.06cm=\hskip -0.06cm 1$-exclusion refers to the unique composition $n=n$, whereas $g\hskip -0.06cm=\hskip -0.06cm 2$-exclusion corresponds to the  standard compositions with no zeros at all. %which were used \cite{nous} for the algebraic area enumeration of  square lattice walks  with  Hofstadter spectral function  (\ref{22}).
 We also   constructed  triangular lattice chiral  walks realizing $g\hskip -0.06cm=\hskip -0.06cm 3$-exclusion 
 with spectral function
 \be{{{{b}}}_{p/q}}(k)=\big(2\sin(2\pi k p/q)\big) \big(2\sin(2\pi (k+1) p/q)\big)\label{44}\ee
 We finally hinted at other walks corresponding to statistics with higher values of the exclusion parameter $g$ and to other spectral functions. 
However, for the triangular lattice chiral walks, as well as for other cases, an explicit algebraic area enumeration formula  was  missing  due to the lack of  binomial expressions  analogous to (\ref{33}) for the  triangular spectral function (\ref{44}).  
 
In the present work  we focus on filling this gap by uncovering such expressions for entire classes of trigonometric 
spectral functions generalizing (\ref{22}) and (\ref{44}). Namely,  we consider, on the one hand
 \be {{{{b}}}_{p/q}}(k)=\big(2\sin(\pi k p/q)\big)^r\label{55}\ee
 and on the other hand
 \be {{{{b}}}_{p/q}}(k) = \big(2\sin(\pi k p/q)\big) \big(2\sin(\pi (k+1) p/q)\big) \ldots \big(2\sin(\pi (k+r-1) p/q)\big)\label{66}\ee
 where in both instances  $r$ can be even or odd. The case $r=2$ reproduces\footnote{The actual spectral function (\ref{44}) for triangular lattice chiral walks has a factor $2$ in front of the $\pi$'s which we omit here to stay in line with (\ref{55}); it anyway amounts to a trivial redefinition of $p/q\to 2p/q$.} (\ref{22}) and (\ref{44}) respectively. We will see that the basic structure of the binomial multiple sum (\ref{33}) naturally generalizes to these cases.  {In the Appendix} we will also derive the relevant generalization for the spectral function
 \be {{{{b}}}_{p/q}}(k)=\big(2\sin(\pi k p/q)\big)^{r/2}\big(2\sin(\pi (k+1) p/q)\big)^{r/2}\label{appendix}\ee
where $r$ is even, yet another possible generalization of (\ref{44}).

Turning  to the algebraic area combinatorics {\it per se}, these expressions, as already mentioned, will allow for explicit enumeration formulae analogous to the square lattice walks formula obtained in \cite{nous} for $g=2$ and the Hofstadter spectral function (\ref{22}). This requires introducing an appropriate  weighting coefficient {in the} summation over compositions of the integer $n$. We refer to \cite{nous}  for detailed explanations of how this procedure unfolds  and to \cite{poly} for the connection  to $g$-exclusion statistics and the resulting generalizations. 
With the $g$-exclusion statistics  weighting coefficients \cite{poly}
\bea
{c_g (l_1,l_2,\ldots,l_{j})} &=& {{{(l_1+\dots +l_{g-1}-1)!\over l_1! \cdots l_{g-1}!}~
\prod_{i=1}^{j-g+1} {l_i+\dots +l_{i+g-1}-1 \choose l_{i+g-1}}}}\nonumber\\
&=& {{{\prod_{i=1}^{j-g+1} (l_i + \dots + l_{i+g-1} -1)! \over \prod_{i=1}^{j-g} (l_{i+1} + \dots +l_{i+g-1} -1 )! } }\prod_{i=1}^j {1\over l_i!} }
\nonumber\eea
we can express the lattice walks algebraic area enumeration for $g\hskip -0.12cm\ge \hskip -0.12cm 2$-exclusion and a general  periodic spectral function
${{{{b}}}_{p/q}}(k)$  by means of the $g$-cluster coefficient\footnote{For statistical mechanics considerations the $g$-cluster coefficient introduced in \cite{poly} is the expression in (\ref{b}) multiplied by $(-1)^{n-1}q/(g n)$.}
\be  b(n)=g n\sbb \sum_{l_1, l_2, \ldots, l_{j}\atop { \text {g-composition}}\;{\rm of}\;n}\sbb\sb c_g(l_1,l_2,\ldots,l_{j} )
\, {1\over q}\sum _{k=1}^q {{{{b}}}_{p/q}}^{{l_1}}(k){{{{b}}}_{p/q}}^{l_2}(k-1)\ldots {{{{b}}}_{p/q}}^{l_{j}}(k-j+1)\label{b}\ee
As already stressed, (\ref{b})  yields the algebraic area combinatorics provided that an expression analogous to (\ref{33}) is known for the specific ${{{b}}_{p/q}}(k)$.  Indeed, the summation index $A$ in (\ref{33}) has to be interpreted in (\ref{b}) as the algebraic area, and the coefficient multiplying the exponential factor $\rme^{\rmi \pi A p/q}$  is the sought for algebraic area counting number. It will, in particular, yield  the triangular lattice chiral walk counting described by $g\hskip -0.06cm= \hskip -0.06cm 3$-exclusion and  spectral function (\ref{44}).

Finally, we will discuss the unexpected occurrence of Ap\'ery-like numbers in the cluster coefficient (\ref{b}) evaluated at particular values of $p/q$ for certain $g$-exclusions and spectral functions. Ap\'ery-like numbers are interesting {\it per se}
since they are cousins of the celebrated Ap\'ery numbers which allow for a proof of the irrationality of $\zeta(2)$ and  $\zeta(3)$. One key characteristic of these numbers is that they are  integer solutions of second order recursion relations. As we will see, some of the $\zeta(2)$ Ap\'ery-like numbers {fascinatingly} emerge in the algebraic enumeration formula (\ref{b}).

%\section{{\bf Trigonometric sums {\scalebox{0.8} {\parbox{\linewidth}{{$\,\sum _{k=1}^q {{{{b}}}_{p/q}}^{\hskip -0.1cm {l_1}}(k)\,{{{{b}}}_{p/q}}^{\hskip -0.1cm{l_2}}(k+1)\,\cdots \,{{{{b}}}_{p/q}}^{\hskip -0.1cm l_{j}}(k+j-1)$}} }}}}

\section{{\bf Trigonometric sums {{$\,\sum _{k=1}^q {{{{b}}}_{p/q}}^{\hskip -0.1cm {l_1}}(k)\,{{{{b}}}_{p/q}}^{\hskip -0.1cm{l_2}}(k+1)\,\cdots \,{{{{b}}}_{p/q}}^{\hskip -0.1cm l_{j}}(k+j-1)$}} }}
 
 We aim at uncovering explicit binomial multiple  sums analogous to (\ref{33}) for the spectral functions (\ref{55}) and (\ref{66}). In fact, the form of (\ref{33}) is quite robust and suggestive, and allows deducing such generalizations by simple deformations while preserving its overall structure.  We stress that, from now on, some $l_i$'s can be null according to the $g$-composition  structure discussed previously, i.e., no more than $g-2$ zeros in succession inside the set. The $A$-summation bounds, when specified, will explicitly depend on the parameter $g$.

  \subsection{Square lattice walks generalization: ${{{{b}}}_{p/q}}(k)=\big(2\sin(\pi k p/q)\big)^r$\label{lean}}
  We first list two basic facts:
\begin{itemize}
\item
 When  $q\to\infty$  one obtains the overall counting 
   \be \int_0^1 \big(2\sin(\pi s)\big)^
   {rl_1+l_2+\ldots+rl_j}\,  ds={r(l_1+l_2+\ldots+l_j)\choose r(l_1+l_2+\ldots+l_j)/2}\label{count}
   \ee
   so  we focus on $(l_1+l_2+\ldots+l_j)$ such that $r(l_1+l_2+\ldots+l_j) $ be even. It means that for $r$ even any set 
   $l_1,l_2,\ldots,l_j$ is admissible, whereas for $r$ odd the $l_i$'s have to be such that their sum be even. 
    \item  It is obvious that for a given $r$ \[{1\over q}\sum _{k=1}^q \big(\big(2\sin(\pi k p/q)\big)^r\big)^{{l_1}}\big(\big(2\sin(\pi (k+1) p/q)\big)^r\big)^{l_2}\ldots \big(\big(2\sin(\pi (k+j-1) p/q)\big)^r\big)^{l_{j}}\]  amounts to 
    \[{1\over q}\sum _{k=1}^q \big(\big(2\sin(\pi k p/q)\big)^2\big)^{rl_1/2}\big(\big(2\sin(\pi (k+1) p/q)\big)^2\big)^{r l_2/2}\ldots \big( \big(2\sin(\pi (k+j-1) p/q)\big)^2\big)^{r l_{j}/2}\]
   which is {essentially} the Hofstadter case $r=2$, i.e., for the  spectral function $\big(2\sin(\pi k p/q)\big)^2$, but now with $l_i\to r l_i/2$.
  \end{itemize}

Based on the above observations, the binomial  multiple sum in (\ref{33}) for the  $r=2$  Hofstadter case becomes, for 
${{{{b}}}_{p/q}}(k)=\big(2\sin(\pi k p/q)\big)^r$  with  $r$ even\footnote{The overall counting, found by replacing $\rme^{ \rmi \pi A p/q}$ by $1$ is
 \begin{align}&{r(l_1+l_2+\ldots+l_j)\choose r(l_1+l_2+\ldots+l_j)/2}=\sum_{A=-(g-1)r\lfloor (l_1+\ldots+l_j)^2/4\rfloor\atop A\;\text{even}}^{(g-1)r\lfloor (l_1+\ldots+l_j)^2/4\rfloor}\nonumber\\\sum_{k_3= -{rl_3/ 2}}^{{rl_3/ 2} 
 }&\ldots \sum_{k_{j}= -{rl_j/ 2}}^{{rl_{j}/ 2}}{rl_1\choose {rl_1/ 2} +A/2+\sum_{i=3}^{j}(i-2)k_i}{rl_2\choose {rl_2/2} -A/2-\sum_{i=3}^{j}(i-1)k_i}\prod_{i=3}^{j}{rl_i\choose {rl_i/ 2}+k_i}\nonumber\end{align}},
\begin{align}
&\sbb{1\over q}\sum _{k=1}^q {{{{b}}}_{p/q}}^{{l_1}}(k){{{{b}}}_{p/q}}^{l_2}(k+1)\ldots {{{{b}}}_{p/q}}^{l_{j}}(k+j-1)=\sum_{A=-(g-1)r\lfloor (l_1+\ldots+l_j)^2/4\rfloor\atop A \;\text{even}}^{(g-1)r\lfloor (l_1+\ldots+l_j)^2/4\rfloor}\rme^{ \rmi \pi A p/q}\label{even}\\
&\sbb\sbb\sum_{k_3= -{rl_3/ 2}}^{{rl_3/ 2} }\ldots \sum_{k_{j}= -{rl_j/ 2}}^{{rl_{j}/ 2}}{rl_1\choose {rl_1/ 2} +A/2+\sum_{i=3}^{j}(i-2)k_i}{rl_2\choose {rl_2/2} -A/2-\sum_{i=3}^{j}(i-1)k_i}\prod_{i=3}^{j}{rl_i\choose {rl_i/ 2}+k_i}
\nonumber\end{align} 
which is valid when $ r(l_1+\ldots+l_j)/2<q$  holds, and where we have specified  the {range} $ [-(g-1)r\lfloor (l_1+\ldots+l_j)^2/4\rfloor,(g-1)r\lfloor (l_1+\ldots+l_j)^2/4\rfloor]$ in which $A$ {needs} to be {restricted}.

  In the $r$ odd case we expect a binomial multiple sum analogous to (\ref{even}). To see this {in} full generality, and to
give a full proof of the original formula with even $r$, let us %rewrite the trigonometric sum (\ref{11}) with spectral function ${{{{b}}}_{p/q}}(k)=\big(2\sin(\pi k p/q)\big)^r$  by 
  first recall the Poisson summation formula
  for any $q$-periodic function $f(x) = f(x+q)$ 
  \be \sum_{k=1}^q f(k)=\sum_{n=-\infty}^{\infty} \tilde{f}(nq)\label{strange}\ee
  where $\tilde{f}$ is the Fourier transform of $f$ defined as
  \be \tilde{f}(k)=\int_0^q f(x)\rme^{- 2\rmi\pi k x/q} dx\;,\;\quad f(x)={1\over q}\sum_{k=-\infty}^{\infty}\tilde{f}(k)\rme^{2\rmi\pi k x/q}
  \nonumber\ee
 Let us  consider the function\ $f(x)={1\over q} {{{{b}}}_{p/q}}^{{l_1}}(x){{{{b}}}_{p/q}}^{l_2}(x+1)\ldots {{{{b}}}_{p/q}}^{l_{j}}(x+j-1)$  which is indeed $q$-periodic due to $r(l_1+l_2+\ldots+l_j)$ being {always}  assumed even. We have 
  \begin{align}  
  \tilde{f}(n q)&=\int_0^q f(k)\rme^{- 2\rmi\pi k n } dk\nonumber\\
  &={1\over q}\int_0^q {{{{b}}}_{p/q}}^{{l_1}}(k){{{{b}}}_{p/q}}^{l_2}(k+1)\ldots {{{{b}}}_{p/q}}^{l_{j}}(k+j-1)\rme^{- 2\rmi\pi k n } dk\nonumber\\
  &={1\over q}\int_0^q \prod_{i=1}^j{1\over \rmi^{rl_i}}\bigg(\rme^{\rmi\pi(k+i-1)p/q}-\rme^{-\rmi\pi(k+i-1)p/q}\bigg)^{rl_i}\rme^{- 2\rmi\pi k n } dk
  \nonumber\\
  &={1\over q}\int_0^q \prod_{i=1}^j{1\over \rmi^{rl_i}}\sum_{k_i=-rl_i/2}^{rl_i/2}{rl_i\choose rl_i/2+k_i}\rme^{2\rmi\pi(k+i-1)k_i p/q}(-1)^{rl_i/2-k_i}\rme^{- 2\rmi\pi k n } dk\nonumber\\&=\sum_{k_1=-rl_1/2}^{rl_1/2}\ldots\sum_{k_j=-rl_j/2}^{rl_j/2}\prod_{i=1}^j{rl_i\choose rl_i/2+k_i}{(-1)^{rl_i/2-k_i}\over \rmi^{rl_i}}\int_0^1 \rme^{2\rmi\pi \sum_{i=1}^j k_i s p+2\rmi\pi\sum_{i=1}^j(i-1)k_i p/q}\rme^{- 2\rmi\pi s q n } ds\nonumber\\
  &=\sum_{k_1=-rl_1/2}^{rl_1/2}\ldots\sum_{k_j=-rl_j/2}^{rl_j/2}\prod_{i=1}^j{rl_i\choose rl_i/2+k_i}{(-1)^{rl_i/2-k_i}\over \rmi^{rl_i}}\rme^{2\rmi\pi\sum_{i=1}^j(i-1)k_i p/q}\,\delta\Big(\sum_{i=1}^jk_i p-n q\Big)
  \label{atreat}\end{align}
As stressed above, {$r(l_1\ldots+l_j)$ is even and thus the sum of the $k_i$ is an integer. Further, $p$ and $q$ are coprime. These facts imply that the Kronecker-$\delta$ in (\ref{atreat})  enforces
  \be p \sum_{i=1}^j k_i = q n ~~~~\text{and thus}~~~~
  \sum_{i=1}^jk_i=t q\quad{\rm and}\quad n=t p\nonumber\ee
  for some integer $t$. Now $\bigl|\sum_{i=1}^jk_i\bigr|\le r(l_1+\ldots+l_j)/2$  and thus, under the condition  $r(l_1+\ldots+l_j)/2<q$, $t$ is necessarily equal to $0$, implying that  $\sum_{i=1}^jk_i=0$ {and $n=0$}.
  From the Poisson summation formula  (\ref{strange}) then we infer
  $\sum_{k=1}^q f(k)=\tilde{f}(0)=\int_0^q f(x) dx\nonumber$;
{that is, for}  ${{{{b}}}_{p/q}}(k)=\big(2\sin(\pi k p/q)\big)^r$,
  \be {1\over q} 
  \sum_{k=1}^q{{{{b}}}_{p/q}}^{{l_1}}(k){{{{b}}}_{p/q}}^{l_2}(k+1)\ldots {{{{b}}}_{p/q}}^{l_{j}}(k+j-1)={1\over q}
  \int_0^q {{{{b}}}_{p/q}}^{{l_1}}(k){{{{b}}}_{p/q}}^{l_2}(k+1)\ldots {{{{b}}}_{p/q}}^{l_{j}}(k+j-1) dk\label{verynice}\ee
   What  has been achieved in (\ref{verynice}) is the trading of
  the  original sum over $k$  from $1$ to $q$ {in the LHS} for the integral over $k$ from $0$ to $q$ {in the  RHS}, which is {valid} provided that $r(l_1+\ldots+l_j)/2<q$. 
  
We can easily check that the trigonometric integral yields  the binomial multiple sum (\ref{33}) in the  $r=2$ case, or  more generally (\ref{even}) in  the $r$  even case.
  To do so let us   proceed  from the last line of (\ref{atreat}):  enforcing the Kronecker $\delta$ in the summand we obtain
\begin{align}  &{1\over q}\int_0^q {{{{b}}}_{p/q}}^{{l_1}}(k){{{{b}}}_{p/q}}^{l_2}(k+1)\ldots {{{{b}}}_{p/q}}^{l_{j}}(k+j-1) dk\nonumber\\&=
{1\over q}\int_0^q
 dk\prod_{i=1}^j 
 \bigg(2\sin\big(\pi k p/q+\pi(i-1)p/q\big)\bigg)^{r l_i}=\int_0^1
 dt\prod_{i=1}^j 
 \bigg(2\sin\big(\pi t+\pi(i-1)p/q\big)\bigg)^{r l_i}\nonumber\\&=\sum_{k_1=-rl_1/2}^{rl_1/2}\ldots\sum_{k_j=-rl_j/2}^{rl_j/2}\prod_{i=1}^j{rl_i\choose rl_i/2+k_i}\rme^{2\rmi\pi\sum_{i=1}^j(i-1)k_i p/q}\label{ro}\end{align}  
The change of integration from $(1/q)\int_0^q dk$ to $\int_0^1 dt$ in the variable $t=k p/q$ 
in the second line  is justified since $r(l_1+\ldots+l_j)$ is even and the integrand has period $1$ in $t$.
We still need to enforce the constraint $\sum_{i=1}^jk_i=0$ in the summation variables $k_i$. To reproduce the $A$-expansion with exponential factors $\rme^{\rmi \pi A p/q}$  in the binomial multiple sums (\ref{33}) and  (\ref{even}), we denote by $A$ the coefficient $2\sum_{i=1}^j(i-1)k_i$ of $\rmi \pi p/q$ appearing in the exponential of the last line in (\ref{ro}). The resulting system of two equations,
$\sum_{i=1}^jk_i=0$  and $A=2\sum_{i=1}^j(i-1)k_i$, can be readily solved for, e.g., the first two variables $k_1$ and $k_2$,
to yield
\be k_1=-A/2+\sum_{i=3}^j(i-2)k_i\quad,\quad k_2=A/2-\sum_{i=3}^j(i-1)k_i\nonumber\ee
Finally, changing summation variables from $k_i$ to $-k_i$ and noting that each binomial is invariant 
under changing the sign of $k_i$, we obtain
\begin{align}
&{1\over q}\sum_{k=0}^q {{{{b}}}_{p/q}}^{{l_1}}(k){{{{b}}}_{p/q}}^{l_2}(k+1)\ldots {{{{b}}}_{p/q}}^{l_{j}}(k+j-1) \nonumber\\
&=\int_0^1 dt\prod_{i=1}^j 
 \bigg(2\sin\big(\pi t+\pi(i-1)p/q\big)\bigg)^{r l_i}\nonumber\\
 &= \sum_{A=-(g-1)r\lfloor (l_1+\ldots+l_j)^2/4\rfloor\atop \text{in steps of 2}}^{(g-1)r\lfloor (l_1+\ldots+l_j)^2/4\rfloor}
 \, \rme^{ \rmi \pi A p/q}\, \sum_{k_3= -{rl_3/ 2}}^{{rl_3/2} 
 }\ldots \sum_{k_{j}= -{rl_j/2}}^{{rl_{j}/ 2}}\label{sonice}\\
&\hskip 0.5cm {rl_1\choose {rl_1/ 2} +A/2+\sum_{i=3}^{j}(i-2)k_i}{rl_2\choose {rl_2/2} -A/2 - \sum_{i=3}^{j}(i-1)k_i}\prod_{i=3}^{j}{rl_i\choose {rl_i/ 2}+k_i}\nonumber
\end{align}
i.e., {precisely} (\ref{even}) {but} now valid for $r$ even and $r$  odd, with a specific $A$-summation dictated by the condition that in (\ref{sonice})  the  first two binomial entries ${rl_1/ 2} +A/2+\sum_{i=3}^{j}(i-2)k_i$ and $ {rl_2/2} -A/2-\sum_{i=3}^{j}(i-1)k_i$  still take integer values  for all   $k_i\in[-rl_i/2,rl_i/2]$, $i=3,\ldots,j$, as was  the case  in (\ref{ro}) for the first two binomial entries ${rl_1/ 2} +k_1$  and $  {rl_2/2} +k_2$ for all  $ k_1\in[-rl_1/2, rl_1/2]$  and $ k_2\in[-rl_2/2, rl_2/2]$. It follows that  in the case $r$ even, where the $k_i$'s are all integers, $A$  has to be even, and in the case $r$ odd, where the $k_i$'s are either integers or half integers, $l_1+l_2+\ldots+l_j$ has to be even  and $ A$ of the same parity as $l_1+l_3+\ldots$ (or  $ l_2+l_4+\ldots$% since  $l_1+l_2+\ldots+l_j$ is even
). In both cases this boils down to $A\in[-(g-1)r\lfloor (l_1+\ldots+l_j)^2/4\rfloor, (g-1)r\lfloor (l_1+\ldots+l_j)^2/4\rfloor\;]$ in steps of $2$. {We also note that, in this and all subsequent formulae, we follow the convention that the sum of all the lower entries in the binomials in (\ref{sonice}) be zero, which fixes the form of such expressions among various
equivalent parametrizations.}
  
We can express the $A$-binomial block in (\ref{sonice}) in an integral form by augmenting the LHS to the double integral ${1\over 2}\int_0^1
 dt\int_0^2
 dt'\prod_{i=1}^j 
 \Big(2\sin(\pi t+\pi(i-1)t'\big)\Big)^{r l_i}\delta(p/q-t')$ and using $2\sum_{n=-\infty}^{\infty}\delta(p/q-t'-2n)=\sum_{A=-\infty}^{\infty}\rme^{\rmi \pi A(p/q-t')}$   to get
 \begin{align}
 &{1\over 2}\int_0^{2}{dt'}\int_0^{1}{dt}\prod_{i=1}^j \bigg(2\sin\big(\pi t+\pi(i-1)t'\big)\bigg)^{r l_i}e^{i\pi At'} \label{top}\\
 = &\hskip -0.2cm \sum_{k_3= -{rl_3/ 2}}^{{rl_3/2} 
 }\hskip -0.1cm\cdots \hskip -0.1cm\sum_{k_{j}= -{rl_j/2}}^{{rl_{j}/ 2}}\sb{rl_1\choose {rl_1/ 2} +\sh A/2+\sum_{i=3}^{j}(i-2)k_i}\sh{rl_2\choose {rl_2/2} -\sb A/2-\sh\sum_{i=3}^{j}(i-1)k_i}\sh\prod_{i=3}^{j}{rl_i\choose {rl_i/ 2}+k_i}\nonumber \end{align}
{In the multiple sum of the RHS $A$ is constrained as above, depending on $r$ being even or odd. However, the integral in the
LHS is valid for all integer values of $A$, yielding zero for the values that do not appear in the RHS.}

 \subsection{Triangular generalization: {${{{{b}}}_{p/q}}(k) = \big(2\sin(\pi k p/q)\big) \big(2\sin(\pi (k+1) p/q)\big)\ldots \big(2\sin(\pi (k+r-1) p/q)\big)$ } }
\vskip -0.1cm
 We can proceed in exactly the same way  for triangular-like spectral functions of the type ${{{{b}}}_{p/q}}(k) = \big(2\sin(\pi k p/q)\big) \big(2\sin(\pi (k+1) p/q)\big) \ldots \big(2\sin(\pi (k+r-1) p/q)\big)$. Again \begin{itemize}
 \item {$q\to\infty$ recovers the overall counting} 
   \be \int_0^1 \big(2\sin(\pi s)\big)^
   {rl_1+rl_2+\ldots+rl_j} ds={r(l_1+l_2+\ldots+l_j)\choose r(l_1+l_2+\ldots+l_j)/2}\nonumber\ee
 as in (\ref{count}), so  we still focus on sets of $l_i$'s  such that $r(l_1+l_2+\ldots+l_j) $ is even, again ensuring the $q$-periodicity of the functions at hand
 \item The rewriting of the trigonometric sum as a trigonometric  integral follows  the same lines as in (\ref{atreat})  under the same condition $r( l_1+\ldots+l_j )/2<q$ since {the sole input in this condition} is
 the highest power of $\rme^{\rmi\pi k p/q}$ that appears in $b_{p/q}(k)$ given by (\ref{66}), which happens to be again    $r$
\end{itemize}

 \subsubsection{Triangular chiral walks $r=2$: ${b_{p/q}}(k)=\big(2\sin(\pi k p/q)\big)\big(2\sin(\pi (k+1) p/q)\big)$ \label{tri2}}
Following the same steps as in {\bf 2.1}, we can rewrite the trigonometric sum corresponding to ${{{{b}}}_{p/q}}(k) = \big(2\sin(\pi k p/q)\big) \big(2\sin(\pi (k+1) p/q)\big)$ as the simple integral
\begin{align} &{1\over q}\sum _{k=1}^q {{{{b}}}_{p/q}}^{{l_1}}(k){{{{b}}}_{p/q}}^{l_2}(k+1)\ldots {{{{b}}}_{p/q}}^{l_{j}}(k+j-1)= \nonumber\\&\int_0^{1}
 {dt}\bigg(2\sin\big(\pi t\big)\bigg)^{ l_1}\prod_{i=2}^j 
 \bigg(2\sin\big(\pi t+\pi(i-1)p/q\big)\bigg)^{l_{i-1}+ l_i}\bigg(2\sin\big(\pi t+\pi j p/q\big)\bigg)^{ l_j}\label{mylady}\end{align}
provided that $l_1+\ldots+l_j<q$. 
 
Integrating (\ref{mylady}) leads to the appropriate  deformation of the binomial multiple sum (\ref{33})  for the spectral function ${{{{b}}}_{p/q}}(k)=\big(2\sin(\pi k p/q)\big)\big(2\sin(\pi (k+1) p/q)\big)$, a deformation  which could also have
been directly guessed by simple manipulations: in (\ref{11}) the integer $l_1$ is associated with the index $k$, $l_1+l_2$ with $k+1$, $l_2+l_3$ with $k+2$, etc. This leads to\footnote {With overall counting, obtained in the $q\to\infty$ limit by replacing $e^{\rmi A p/q}$ by 1:
   \be\nonumber
   {2(l_1+l_2+\ldots+l_j)\choose l_1+l_2+\ldots+l_j}\ee
} 
\begin{align} &{1\over q}\sum _{k=1}^q {{{{b}}}_{p/q}}^{{l_1}}(k){{{{b}}}_{p/q}}^{l_2}(k+1)\ldots {{{{b}}}_{p/q}}^{l_{j}}(k+j-1)\nonumber\\&=\int_0^{1}
 {dt}\bigg(2\sin\big(\pi t\big)\bigg)^{ l_1}\prod_{i=2}^j 
 \bigg(2\sin\big(\pi t+\pi(i-1)p/q\big)\bigg)^{l_{i-1}+ l_i}\bigg(2\sin\big(\pi t+\pi j p/q\big)\bigg)^{ l_j}\nonumber\\&=\sum_{A=-\lceil(l_1+\ldots+l_j)^2/2\rceil-(g-2)\lfloor(l_1+\ldots+l_j)^2/2\rfloor\atop A\;\text{same parity as}\; l_1+l_2+\ldots+l_j}^{\lceil(l_1+\ldots+l_j)^2/2\rceil+(g-2)\lfloor(l_1+\ldots+l_j)^2/2\rfloor}\rme^{ \rmi \pi A p/q}\sum_{k_3=-(l_2+l_3)/2}^{(l_2+l_3)/2} 
\ldots \sum_{k_j = -(l_{j-1}+l_j)/2}^{( l_{j-1}+l_j)/2}\sum_{k_{j+1}= -l_j/2}^{l_{j}/2}\nonumber\\&{l_1\choose {l_1/2 +A/2+\sum_{i=3}^{j+1}(i-2)k_i}}{l_1+l_2\choose (l_1+l_2)/2 -A/2-\sum_{i=3}^{j+1}(i-1)k_i}\nonumber\\&\times\prod_{i=3}^{j}{l_{i-1}+l_i\choose (l_{i-1}+l_i)/2+k_i}{l_j\choose l_j/2+k_{j+1}}\label{triangular}\end{align} 

{We note that $A$ in the summation (\ref{triangular}) spans the interval  $[-\lceil(l_1+\ldots+l_j)^2/2\rceil-(g-2)\lfloor(l_1+\ldots+l_j)^2/2\rfloor, \lceil(l_1+\ldots+l_j)^2/2\rceil+(g-2)\lfloor(l_1+\ldots+l_j)^2/2\rfloor\;]$ increasing by steps of $2$}, which in particular implies that $A$ is of the same parity as $l_1+l_2+\ldots+l_j$.

\subsubsection{$r=3$: ${{{{b}}}_{p/q}}(k)=\big(2\sin(\pi k p/q)\big)\big(2\sin(\pi (k+1) p/q)\big)\big(2\sin(\pi (k+2) p/q)\big)$  \\ with $l_1+\ldots+l_j$ even \label{tri3}}

Similarly to the previous cases one can rewrite  the $r=3$ triangular trigonometric   sum   as the simple integral
\begin{align} &{1\over q}\sum _{k=1}^q {{{{b}}}_{p/q}}^{{l_1}}(k){{{{b}}}_{p/q}}^{l_2}(k+1)\ldots {{{{b}}}_{p/q}}^{l_{j}}(k+j-1)= \nonumber\\&\int_0^{1}
 {dt}\bigg(2\sin\big(\pi t\big)\bigg)^{ l_1}\bigg(2\sin\big(\pi t+\pi p/q\big)\bigg)^{l_{1}+ l_2}\prod_{i=3}^j 
 \bigg(2\sin\big(\pi t+\pi(i-1)p/q\big)\bigg)^{l_{i-2}+l_{i-1}+ l_i}\nonumber\\&\times \bigg(2\sin\big(\pi t+\pi j p/q\big)\bigg)^{ l_{j-1}+l_j}\bigg(2\sin\big(\pi t+\pi (j+1) p/q\big)\bigg)^{l_j}\nonumber\end{align}
 provided that $3(l_1+\ldots+l_j)/2<q$.

 Likewise one obtains the binomial multiple sum\footnote{With overall counting, obtained by replacing $e^{\rmi\pi A p/q}$ by 1:
   \be\nonumber
   {3(l_1+l_2+\ldots+l_j)\choose 3(l_1+l_2+\ldots+l_j)/2}\ee
}
 \begin{align} &{1\over q}\sum _{k=1}^q {{{{b}}}_{p/q}}^{{l_1}}(k){{{{b}}}_{p/q}}^{l_2}(k+1)\ldots {{{{b}}}_{p/q}}^{l_{j}}(k+j-1)=\sum_{A=-\infty\atop A\;\text{same parity as}\; l_1 +l_3+\ldots\;{\it or}\; l_2+l_4+\ldots}^{\infty}\rme^{\rmi\pi A p/q}\nonumber\\&\sum_{k_3=-(l_1+l_2+l_3)/2}^{(l_1+l_2+l_3)/2} 
\ldots \sum_{k_j = -(l_{j-2}+l_{j-1}+l_j)/2}^{(l_{j-2}+ l_{j-1}+l_j)/2}\;\;\sum_{k_{j+1}= -(l_{j-1}+l_j)/2}^{(l_{j-1}+l_{j})/2}\;\;
\sum_{k_{j+2}= -l_j/2}^{l_{j}/2}\nonumber\\&\hskip 0.2cm{l_1\choose l_1/2 +A/2+\sum_{i=3}^{j+2}(i-2)k_i}{l_1+l_2\choose (l_1+l_2)/2 -A/2-\sum_{i=3}^{j+2}(i-1)k_i}\nonumber\\&\hskip 0.2cm\times\prod_{i=3}^{j}{l_{i-2}+l_{i-1}+l_i\choose (l_{i-2}+l_{i-1}+l_i)/2+k_i}{l_{j-1}+l_j\choose (l_{j-1}+l_j)/2+k_{j+1}}{l_j\choose l_j/2+k_{j+2}}\label{triangr3}\end{align} 
where  $A$  has to be of the same parity as $l_1+l_3+\ldots $ (or $l_2+l_4+\ldots $) {and obviously a finite range. }{The cases $r=4$ and beyond are treated in the Appendix.}

\section{Algebraic area enumeration and Ap\'ery-like \\ numbers  }

\subsection{\bf Algebraic area  enumeration}

We can retrieve from the cluster coefficient (\ref{b}) algebraic area enumeration formulae for various random lattice walks. % In \cite{poly} we focused on $q$-periodic spectral functions so that the $g$-exclusion statistical mechanics setting is operative. %From now on we restrict to such periodic functions which means $r$ even (when $r$ is odd the spectral functions are anti-periodic).
For example, from (\ref{sonice}) for ${{{{b}}}_{p/q}}(k)=\big(2\sin(\pi k p/q)\big)^r$  with $r$ even and  $g$-exclusion,  (\ref{b}) becomes
\begin{align} & b(n)= g n  \sum_{A=-(g-1)r\lfloor (l_1+\ldots+l_j)^2/4\rfloor\atop A\;\text{even}}^{(g-1)r\lfloor (l_1+\ldots+l_j)^2/4\rfloor}\rme^{ \rmi \pi A p/q} \sbb \sum_{l_1, l_2, \ldots, l_{j}\atop g\text{-composition of}\; n} c_g (l_1,l_2,\ldots,l_{j}) \label{reven}\\&\sbb\sbb\sum_{k_3= -{rl_3/ 2}}^{{rl_3/2} 
 }\ldots \sum_{k_{j}= -{rl_j/2}}^{{rl_{j}/ 2}}\sb{rl_1\choose {rl_1/ 2} +\sb A/2+\sum_{i=3}^{j}(i-2)k_i}\sh{rl_2\choose {rl_2/2} -\sb A/2-\sum_{i=3}^{j}(i-1)k_i}\sh\prod_{i=3}^{j}{rl_i\choose {rl_i/ 2}+k_i}\nonumber\end{align}
 with overall counting, given by replacing   $\rme^{ \rmi \pi A p/q}$ by $1$
 \bea && {gn\choose n}{rn\choose rn/2}\label{countingbis}\eea
 The second binomial in (\ref{countingbis}), as initially discussed in (\ref{333}) and  displayed  in the various overall counting cases of  subsection (\ref{lean}), results from the trigonometric sums replacing  $\rme^{ \rmi \pi A p/q}$ by $1$  in the limit $q\to\infty$,  whereas the first one results from  the summation of the exclusion weight coefficients $c_g$  over all $g$-compositions of the integer $n$. }  

\subsubsection{Square lattice walks: ${{{{b}}}_{p/q}}(k)=\big(2\sin(\pi k p/q)\big)^2$ \label{myHof}}
 
As already {stated},  the standard square lattice walks are specifically $g=2$  and $r=2$ and are defined in terms of the Hamiltonian \cite{poly} \be H=(1-u)v+v^{-1}(1-u^{-1})\nonumber\ee where $u$ and $v$ respectively stand for the {\it right} and {\it up}  hopping operators on the  lattice, with commutation $vu=\scq \, uv$, where $\scq=e^{\rmi \Phi} = e^{\rmi 2\pi p/q}$ is the noncommutativity parameter encoding the presence of the magnetic field perpendicular to the lattice, with $\Phi$ the magnetic flux per plaquette. We recover the Hofstadter spectral function as
\be b_{p/q}(k)=(1-\scq^{-k})(1-\scq^{k})=\big(2\sin(\pi k p/q)\big)^2
\nonumber\ee

The Hamiltonian  describes a random walk with elementary steps {\it up},   {\it right} followed by {\it up}, {\it down}, and {\it down}  followed by  {\it left}. It means that starting  from the origin $(0,0)$ it reaches after one step the lattice points $(0,1)$, $(1,1)$, $(0,-1)$ or $(-1,-1)$ with equal probability. This generates deformed walks on the square lattice (see Fig.1) which are equivalent through a modular transformation to the usual square lattice walks.
(This modular transformation amounts to the transformation $u \to - u v$, which leaves the $u,v$ commutation relation unchanged and turns $H$ into $u+v+u^{-1}+v^{-1}$.)
$b(n)$ in  (\ref{reven}) then yields the desired algebraic area counting  \cite{nous}
\[b(n)=\sum_{A=-2\lfloor n^2/4\rfloor\atop A\;\text{even}}^{2\lfloor n^2/4\rfloor}  \rme^{ \rmi \pi A p/q}C_{2n}(A)\] 
where \bea && C_{2n}(A)= 2 n   \sum_{l_1, l_2, \ldots, l_{j}\atop\text{2-composition of}\; n} c_2 (l_1,l_2,\ldots,l_{j})\label{toto2}\\&&\sum_{k_3= -{l_3}}^{{l_3} 
 }\ldots \sum_{k_{j}= -{l_j}}^{{l_{j}}}{2l_1\choose {l_1} +A/2+\sum_{i=3}^{j}(i-2)k_i}{2l_2\choose {l_2} -A/2-\sum_{i=3}^{j}(i-1)k_i}\prod_{i=3}^{j}{2l_i\choose {l_i}+k_i}\nonumber\eea 
 with $A$ even in the interval $[-2\lfloor n^2/4\rfloor,2\lfloor n^2/4\rfloor]$. $C_{2n}(A)$ 
 counts the number of closed square lattice walks   of length $2n$ --there are overall ${2n\choose n}^2$ of them, see (\ref{countingbis})-- enclosing  an algebraic area  
 %$A/2$ with $-[(2n)^2/16] \le A/2 \le [(2n)^2/16]$ i.e., $-\left \lfloor{n^2/4}\right \rfloor\le A/2 \le\left \lfloor{n^2/4}\right \rfloor $.
{$A/2$} in the interval\footnote{This can be easily seen geometrically for lattice walks of length $2n$ with $n$ even, which have largest possible area $\pm (n/2)^2$: this is the walk circling a square of side $n/2$ anti-clockwise or clockwise.}  $[-\lfloor{n^2/4} \rfloor, \lfloor{n^2/4} \rfloor\;]$: indeed the mapping of {random walk algebraic area to the Hofstadter model} \cite{nous} is via {the weighting factor} $\scq^\text{algebraic area}$,  where $\scq=e^{2i\pi p/q}$,  so here, with $e^{i\pi Ap/q}$ appearing in (\ref{reven}), the algebraic area is  $A/2$.
 
 \subsubsection{Square lattice walks: ${b_{p/q}}(k)=\big(2\sin(\pi k p/q)\big)^4$ \label{myHofbis}}
 
Let us now look at  square lattice walks with $g=2$  and $r=4$ which are defined in terms of the Hamiltonian 
\be H=(u+u^{-1})^2 v+v^{-1} (u+u^{-1})^2\label{oups}\ee
 The corresponding spectral function 
\be b_{p/q}(k)=(\scq^k+\scq^{-k})^4=\big(2\cos(2\pi k p/q)\big)^4
\nonumber\ee
can be put in the standard form (\ref{55}) for $r=4$  by redefining $u\to \rmi u$ and $\scq\to \sqrt{\scq}$,  which does not affect the counting of walks nor the area weighting.

The Hamiltonian (\ref{oups}) describes a random walk with elementary steps  in groups of one random step {\it up} or {\it down} and two independent random steps {\it right} or {\it left}. It means that starting  from the origin $(0,0)$ it reaches after one step the lattice points $(2,1)$, $(-2,1)$, $(2,-1)$ or $(-2,-1)$  with probability $1/8$, or the lattice points  $(1,0)$ or $(-1,0)$ with probability $1/4$. {The same walk can be described as a particle hopping on an {\it even} or {\it odd} square sublattice, where even points are those with $x$ and $y$ coordinates adding to an even integer, the remaining being odd.
The walk proceeds randomly on one of the sublattices but at each step it has the option to move to the nearest up or down point of the opposite sublattice, with each such jump contributing a factor of two in the weight of the walk. The Hamiltonian (\ref{oups}) counts the weighted number of such closed walks of a given total area.}

There are ${2n\choose n}{4n\choose 2n}$  such closed walks of length $2n$, as in (\ref{countingbis}).
The enumeration of such walks enclosing a given algebraic area, with the proper weight, is given by (\ref{reven}): 
\[b(n)=\sum_{A=-4\lfloor n^2/4\rfloor\atop A\;{\it even}}^{4\lfloor n^2/4\rfloor}  \rme^{ \rmi \pi A p/q}C'_{2n}(A)\] 
 where
 \bea && C'_{2n}(A)= \; 2 n   \sbb\sbb\sum_{l_1, l_2, \ldots, l_{j}\atop\text{2-composition of}\; n} c_2 (l_1,l_2,\ldots,l_{j}) \nonumber\\&&\sum_{k_3= -{2l_3}}^{{2l_3} 
 }\ldots \sum_{k_{j}= -{2l_j}}^{{2l_{j}}}{4l_1\choose {2l_1} \sb+\sb A/2+\sum_{i=3}^{j}(i-2)k_i}{4l_2\choose {2l_2}
 \hskip -0.05cm -\sb A/2-\sum_{i=3}^{j}(i-1)k_i}\prod_{i=3}^{j}{4l_i\choose {2l_i}+k_i}\nonumber
 \label{toto4}\eea 
 with $A$ even in the interval $[-4\lfloor n^2/4\rfloor,4\lfloor n^2/4\rfloor]$. $C'_{2n}(A)$ 
 counts the number of closed square lattice walks described above  of length $2n$ and enclosing  an algebraic area  
 %$A/2$ with $-[(2n)^2/16] \le A/2 \le [(2n)^2/16]$ i.e., $-\left \lfloor{n^2/4}\right \rfloor\le A/2 \le\left \lfloor{n^2/4}\right \rfloor $.
{$A/2$}.
% in the interval  $[-2\lfloor{n^2/4} \rfloor, 2\lfloor{n^2/4} \rfloor\;]$.
\vskip -0.2cm
\subsubsection{Square lattice walks: ${b_{p/q}}(k)=\bigg(\big(2\sin(\pi k p/q)\big)\big(2\sin(\pi (k+1) p/q)\big)\bigg)^2$\label{myHofter}}
 
Now consider square lattice walks with $g=2$  and $r=4$ defined by the Hamiltonian 
\be H=(u+u^{-1})v(u+u^{-1})+(u+u^{-1})v^{-1}(u+u^{-1})\label{oupsbis}\ee

 The  spectral function can be brought to the standard form (\ref{appendix}) for $r=4$ by an appropriate redefinition of $u \to -\rmi u$
 \be {b_{p/q}}(k)=\big(2\sin(\pi k p/q)\big)^2 \big(2\sin(\pi (k+1) p/q)\big)^2
\nonumber\ee
Its treatment is given in the subsection \ref{nono0} of the Appendix.

This walk proceeds with sets of one step left or right, one step up or down and another step left or right.
With an appropriate redefinition of $u$ and $v$ (modular transformation) this
walk can also be mapped to a walk proceeding on odd or even square sublattices, as in the last subsection, but now the weight
of jumping on the opposite sublattice is not 2, as before, but rather $Q + Q^{-1}$. So in this description 
the weight of the walks depends explicitly on $Q$, unlike any other walk we encountered before.

There are again ${2n\choose n}{4n\choose 2n}$ such closed walks of length $2n$.% in the limit $q\to\infty$ (i.e., $Q\to 1$). 
The enumeration of such walks enclosing a given algebraic area, with the proper weight, is given by
\[b(n)=\sum_{A=-\infty\atop A\;\text{even}}^{\infty}  \rme^{ \rmi \pi A p/q}C''_{2n}(A)\] 
 where
 \begin{align} &C''_{2n}(A)= \;2 n   \sbb\sum_{l_1, l_2, \ldots, l_{j}\atop\text{2-composition of}\; n} c_2 (l_1,l_2,\ldots,l_{j})
 \nonumber \\
 &\sbb\sum_{k_3=-(l_2+l_3)}^{l_2+l_3} \sb
\ldots \sb \sum_{k_j = -(l_{j-1}+l_j)}^{ l_{j-1}+l_j)}\sum_{k_{j+1}= -l_j}^{l_{j}} {2l_1\choose {l_1 \sb +\sb A/2+\sum_{i=3}^{j+1}(i-2)k_i}}{2(l_1+l_2)\choose l_1\sb +\hskip -0.05cm l_2\hskip -0.05cm -\sb A/2-\sum_{i=3}^{j+1}(i-1)k_i}\nonumber\\&\hskip 5.3cm \times\prod_{i=3}^{j}{2(l_{i-1}+l_i)\choose l_{i-1}+l_i+k_i}{2l_j\choose l_j+k_{j+1}}\label{zizi4}\end{align} 
$C''_{2n}(A)$ counts again the weighted number of closed square lattice walks described above of length $2n$ enclosing  an algebraic area {$A/2$}. It differs from the corresponding number (\ref{toto4}) only in the weighting factor when jumping sublattices.

 %$A/2$ with $-[(2n)^2/16] \le A/2 \le [(2n)^2/16]$ i.e., $-\left \lfloor{n^2/4}\right \rfloor\le A/2 \le\left \lfloor{n^2/4}\right \rfloor $.

  \subsubsection{Triangular lattice chiral walks: ${{{{b}}}_{p/q}}(k) = \big(2\sin(\pi k p/q)\big) \big(2\sin(\pi (k+1) p/q)\big)$  \label{krew} }
From (\ref{triangular}) for the  triangular spectral function (\ref{66}) with $r=2$ and $g$-exclusion we obtain
\bea && b(n)=g n \sum_{A=-\lceil n^2/2\rceil-(g-2)\lfloor n^2/2\rfloor\atop A\;\text{same parity as}\; n}^{\lceil n^2/2\rceil+(g-2)\lfloor n^2/2\rfloor} \rme^{\rmi\pi A p/q} \sum_{l_1, l_2, \ldots, l_{j}\atop\text{g-composition of}\; n} c_{g} (l_1,l_2,\ldots,l_{j})\nonumber\\\nonumber&&\sum_{k_3=-(l_2+l_3)/2}^{(l_2+l_3)/2} 
\ldots \sum_{k_j = -(l_{j-1}+l_j)/2}^{( l_{j-1}+l_j)/2}\sum_{k_{j+1}= -l_j/2}^{l_{j}/2}\nonumber\\&&{l_1\choose {l_1/2 +A/2+\sum_{i=3}^{j+1}(i-2)k_i}}{l_1+l_2\choose (l_1+l_2)/2 -A/2-\sum_{i=3}^{j+1}(i-1)k_i}\nonumber\\&&\times \prod_{i=3}^{j}{l_{i-1}+l_i\choose (l_{i-1}+l_i)/2+k_i}{l_j\choose l_j/2+k_{j+1}}\label{toto}\eea
with overall counting {given} by replacing $\rme^{\rmi\pi A p/q}$ by 1 
\[{gn\choose n}{2n\choose n}\]
Triangular {$g=3$} lattice chiral walks  correspond to the quantum Hamiltonian
\be\nonumber
H = i(- u+ u^{-1})\, v +  v^{-2}
\ee
with  spectral function 
\be
{{{b}}}_{p/q}(k) =\Big(2 \sin {2 \pi p k \over q}\Big)\Big(2\sin{2 \pi p (k+1) \over q}\Big)\nonumber
\ee
as already given in (\ref{44}).
They are  depicted in Figs.2--4  (see \cite{poly} for more details; these walks are the generalization to four quadrants  of the Kreweras walks \cite{kreweras}). 
Since the exclusion parameter is $g=3$  the counting above reduces to
\[{3n\choose n,n,n}\]  which is the number of closed triangular lattice chiral walks of length $3n$. 
The cluster coefficient  (\ref{toto}) then yields the triangular lattice chiral walks algebraic area counting
\[ b(n)= \sum_{A=-n^2\atop A\;\text{in steps of}\; 2}^{n^2}  \rme^{ \rmi \pi A p/q}C_{3n}(A)\]
where
\bea C_{3n}(A)=&& 3 n  \sbb\sum_{l_1, l_2, \ldots, l_{j}\atop\text{3-compositions of}\; n} c_{3} (l_1,l_2,\ldots,l_{j})\sum_{k_3=-(l_2+l_3)/2}^{(l_2+l_3)/2} 
\ldots \sum_{k_j = -(l_{j-1}+l_j)/2}^{( l_{j-1}+l_j)/2}\sum_{k_{j+1}= -l_j/2}^{l_{j}/2} \nonumber\\&&{l_1\choose {l_1/2 +A/2+\sum_{i=3}^{j+1}(i-2)k_i}}{l_1+l_2\choose (l_1+l_2)/2 -A/2-\sum_{i=3}^{j+1}(i-1)k_i}\nonumber\\&&\times \prod_{i=3}^{j}{l_{i-1}+l_i\choose (l_{i-1}+l_i)/2+k_i}{l_j\choose l_j/2+k_{j+1}}\label{totobis}\eea with $A$ in the interval $[-n^2, n^2]$ with same parity as $n$.

$C_{3n}(A)$  counts the number of closed triangular lattice  chiral  walks of length $3n$ enclosing  an algebraic area $A$. Indeed, the mapping of triangular algebraic area-quantum triangular Hamiltonian discussed in \cite{poly} is via $\scq^\text{algebraic area}$ where $\scq =e^{2\rmi\pi p/q}$.
Since in $b_{p/q}(k)$ of (\ref{66}) the building block  $2\sin(\pi k p/q)$ 
is used, rather than $2\sin(2\pi k p/q)$ as in (\ref{44}), we end up with  $e^{\rmi\pi A p/q}$ in (\ref{toto}) in place of
$e^{2\rmi\pi A p/q}$, so that the algebraic area is $A$. One can directly check by explicit enumeration that when $n$ is odd $A$ is also odd
(see, e.g., $n=1$ with $3$ walks of algebraic area $1$ and $3$ walks of algebraic area $-1$) and when $n$ is even $A$ is 
also even (as in $n=2$, with algebraic areas $0, \pm 2$ and $\pm 4$).

 We conclude our discussion of algebraic area counting by remarking that it was possible to extract explicit expressions in terms
of binomial sums for $C_{2n}(A)$ in (\ref{toto2}), $C'_{2n}(A)$ in (\ref{toto4}) and $C_{3n}(A)$ in (\ref{totobis}) from the cluster coefficients  (\ref{reven}) or (\ref{toto}) because the summation constraints over $A$ in the relevant binomial multiple  sums (\ref{sonice}) with $r=2,4$ ($A$ even) or (\ref{triangular}) with $r=2$ ($A$ same parity as $n$), as well as the summation ranges, depend only on $n$ and not on the $l_i$'s themselves.
Similar expressions would apply for walks deriving from odd $r$ binomial sums, like (\ref{sonice}) or (\ref{triangr3}), provided that the binomials appearing in the expressions are understood to vanish for values of $A$ leading to noninteger
entries, as discussed after (\ref{sonice}).

It is a curious fact that if, in the binomial  multiple sums or the cluster coefficients, we sum over {\it all} integer values of $A$ without
restrictions, and analytically continue the binomials to fractional values using Gamma functions, the resulting infinite
sums are closely related to the finite ones over the allowed values of $A$. This point is detailed and explained in the subsection  \ref{nono} of the Appendix.
%It means,  considering  for example the binomial multiple sum (\ref{sonice}), that for even $r$ and any set of $l_i$'s, the infinite sequence of coefficients of {\it odd} $A$, which are rational numbers times $1/\pi^2$, has cumulative sums converging to the standard binomial counting ${r(l_1+l_2+\ldots+l_j)\choose r(l_1+l_2+\ldots+l_j)/2}$.
It means,  considering  for example the binomial multiple sum (\ref{sonice}), that for even $r$ and any set of $l_i$'s, the cumulative sum of the infinite sequence of coefficients of {\it odd} $A$, which are rational numbers times $1/\pi^2$, 
converges to the standard binomial counting ${r(l_1+l_2+\ldots+l_j)\choose r(l_1+l_2+\ldots+l_j)/2}$.

 \subsection{Ap\'ery-like numbers}
 We finally turn to the occurrence of Ap\'ery-like numbers in cluster coefficients (\ref{b}) when evaluated at certain values of $p/q$. We stress that we no more view $b(n)$ as generating algebraic area enumerations of actual lattice walks, but instead consider it as a stand-alone mathematical entity that happens to lead to such occurrences.
 
 % In this context there is no need anymore to restrict to $q$-periodic spectral functions.

 \subsubsection{ Ap\'ery-like numbers  $g=2$  and $r=2$ : ${{{{b}}}_{p/q}}(k)=\big(2\sin(\pi k p/q)\big)^2$ \label{Hof}}
Let us consider\footnote{Or equivalently, using (\ref{verynice})
\bea && b(n)= g n \sum_{l_1, l_2, \ldots, l_{j}\atop{\it g-composition}\;{\it of}\; n} c_g (l_1,l_2,\ldots,l_{j}) \int_0^{1}
 {dt}\prod_{i=1}^j 
 \bigg(2\sin\big(\pi t+\pi(i-1)p/q\big)\bigg)^{r l_i}\nonumber\eea}  $b(n)$ in (\ref{reven}).
For $g=2$ and $r=2$ it gives, for $n=1,2,3,\ldots$  
 \bea p/q=1\quad \Rightarrow && b(n)={2n\choose n}^2 \Leftrightarrow  {\rm closed}\;{\rm square}\;{\rm lattice}\;{\rm walks}\;{\rm counting}\nonumber\\
   p/q=1/2 \Rightarrow && b(n)= 4, 20, 112, 676, 4304, 28496, \ldots\nonumber\eea
    These are the Ap\'ery-like numbers $\zeta(2)$ sequence OEIS A081085   
   \be \sum_{k= 0}^{
   n}{n\choose k} {2 k\choose k} {2 n - 2k\choose n - k}= \sum_{k= 0}^{
   [n/2]}4^{n-2k}{n\choose 2k} {2 k\choose k}^2 \nonumber\ee
with recurrence relation 
   \be (n + 1)^2 b(n + 1) - \big(12 n( n+1) + 4\big) b(n) + 32 n^2 b(n - 1)=0\nonumber\ee
   
   \subsubsection{Ap\'ery-like numbers  $g=2$ and $r=1$ : ${{{{b}}}_{p/q}}(k)=2\sin(\pi k p/q)$   }
Let us still focus on (\ref{reven}) but now for $g=2$ and $r=1$, with $n$ necessarily even\footnote{$n$ is necesseraly even  because $l_1+l_2+\ldots+\l_j$ (which is equal to $n$) has to be even.}. 
We find, for $n=2,4,6,\ldots$ 
\bea p/q=1\quad \Rightarrow && b(n)=(-1)^{n/2}{n\choose n/2}^2 \nonumber\\
   p/q=1/2 \Rightarrow && b(n)= 4, 20, 112, 676, 4304, 28496, \ldots\nonumber\eea
These are the same Ap\'ery-like numbers as above 
   \be \sum_{k= 0}^{
   n/2}{n/2\choose k} {2 k\choose k} {n - 2k\choose n/2 - k} \nonumber\ee
   now occurring for even $n$'s.
   \noindent Indeed, cases $r=2$ and ($r=1$, $n$ even) are essentially equivalent: calling $n=2n'$ for $r=1$, 
  then $\big(2\sin(\pi kp/q)\big)^{n=l_1+l_2+\ldots +l_j}$ with $l_1 ,l_2, \ldots, l_j$ a composition of $n$, is in fact 
  $ \big(\left(2\sin(\pi kp/q)\right)^2\big)^{l'_1+l'_2+\ldots +l'_j=n'}$  with $l'_1 ,l'_2, \ldots, l'_j$ a composition of $n'$, which
  is the $ r=2$ result. 
      
   {\subsubsection{ Ap\'ery-like numbers   $g=2$  and $r=4$: {${{{{b}}}_{p/q}}(k)=\big(2\sin(\pi k p/q)\big)^4$ } }}
Let us again focus on $b(n)$ in (\ref{reven}) but now for $g=2$  and $r=4$: we find, for $n=1,2,3,\ldots$ 
   \bea p/q=1\quad \Rightarrow && b(n)={2n\choose n}{4n\choose 2n}\nonumber\\
    p/q=1/2\Rightarrow && b(n)=12, 164, 2352, 34596, 516912, 7806224,\ldots\nonumber\eea
  These are the Ap\'ery-like numbers $\zeta(2)$ sequence OEIS A143583 
   \be \sum_{k= 0}^{
   n}{2 k\choose k} {4 k\choose 2 k}{2 n - 2 k\choose 
     n - k}{4 n - 4 k\choose 2 n - 2 k}/{2 n\choose  n} =\sum_{k= 0}^{
   n}4^{n-k}{2n-2k\choose n-k} {2 k\choose k}^2\nonumber\ee
   with  recurrence relation
  \be (n + 1)^2 b(n + 1) - (32 n( n+1) + 12) b(n) + 256 n^2 b(n - 1)=0\nonumber\ee

  \subsubsection{ Ap\'ery-like numbers  $g=3$  and $r=2$:{${{{{b}}}_{p/q}}(k) = \big(2\sin(\pi k p/q)\big) \big(2\sin(\pi (k+1) p/q)\big)$}}
Finally we focus \footnote{Or equivalently, using  (\ref{mylady}), on
\bea  b(n)=&&g n \sum_{l_1, l_2, \ldots, l_{j}\atop\text{g-composition of}\; n} c_g (l_1,l_2,\ldots,l_{j})\nonumber\\&&\int_0^{1}
 {dt}\bigg(2\sin\big(\pi t\big)\bigg)^{ l_1}\prod_{i=2}^j
 \bigg(2\sin\big(\pi t+\pi(i-1)p/q\big)\bigg)^{l_{i-1}+ l_i}\bigg(2\sin\big(\pi t+\pi j p/q\big)\bigg)^{ l_j}\nonumber\eea}
 on $b(n)$ in (\ref{toto}).
We find, for $g=3$ and $n=1,2,3,\ldots$
\bea p/q=1\quad\Rightarrow b(n)&=& (-1)^n{3n\choose n}{2n\choose n}\Leftrightarrow {\rm triangular}\;{\rm lattice}\;{\rm chiral}\;{\rm walks}\;{\rm counting}\nonumber\\
\nonumber p/q=1/2\Rightarrow b(n)&=& {3n/2\choose n/2}{n\choose n/2} \nonumber\quad{\rm if}\; n\;{\rm multiple}\;{\rm of}\;2 \;{\rm and}\; 0\;{\rm otherwise}\\
    p/q=1/3\Rightarrow b(n)&=& 3, 9, 21, 9, -297, -2421,\ldots\nonumber
     \eea
     These are Ap\'ery-like numbers $\zeta(2)$ sequence OEIS A006077
\[
 \sum_{k= 0}^{[n/3]}(-1)^k 3^{n - 3 k}{n\choose 
    3 k}{2 k\choose k} {3k \choose k} = \sum_{k= 0}^{[n/3]}(-1)^k 3^{n - 3 k}{n\choose 
    n-3 k, k, k,k}\]
    with recurrence relation
\[(n + 1)^2 b(n + 1) + \big(9 n (n + 1) + 3\big) b(
    n) + 27 n^2 b(n - 1)=0\]
    
   \section{Conclusions}

The trigonometric identities analyzed in this work, as well as their generalizations to other spectral functions that can be derived
along the lines presented here, allow us to obtain expressions for the algebraic area counting of a broad set of random
walks on two-dimensional lattices. The only requirement is that these walks be described by a Hamiltonian of the general
form introduced in \cite{poly}, admitting an interpretation as systems of generalized exclusion statistics with specific spectral functions. A wide class of lattice walk models can be embedded into this framework, and we gave a few examples in the
present work, most notably the triangular chiral walk introduced originally in \cite{poly}.

The most obvious and interesting extension of our results would be in obtaining the area counting of other, more general types
of walks. From the algebraic point of view, an immediate choice presents itself: the Hamiltonian
\be
H_m =(u+u^{-1})^m\, v+v^{-1}(u+u^{-1})^m ~,~~~ m = 1,2,\dots
\nonumber \ee
describes a class of Hofstadter-like models representing generalized random walks on the square lattice, with $m=1$ the
standard (Hofstadter) random walk and $m=2$ the walk studied in subsection {\bf \ref{myHofbis}}.
The model for general
$m$ represents a walk that proceeds in groups of one random step up or down and then $m$ independent random
steps left or right, but other representations are possible by performing modular transformations to the lattice
(or redefinitions of the $u,v$ operators in the Hamiltonian).
All these walks belong to the class of $g=2$ exclusion statistics and their area counting is readily given by the relevant
$g=2$ cluster coefficients and generalized trigonometric sums.

Clearly this is just the tip of a large iceberg as far as lattice walk models are concerned. For instance, another class of walks
at $g=2$ would be described by the Hamiltonian
\be
{\tilde H}_m = (u^m + u^{m-1} + \cdots + u^{-m} ) v + v^{-1} (u^m + u^{m-1} + \cdots + u^{-m} )
\nonumber\ee
This represents walks proceeding with a random step up or down to one of the $2m+1$ neighboring points in the left-right
direction of distance up to $m$ from the original horizontal position with equal probability. Again, the combinatorics of these
walks are readily obtained with our methods. Yet other walks can be constructed, with asymmetrical propagation rules
and belonging to higher $g$ statistics. The only limitation, or criterion, is the potential relevance and physical significance
of these walks, and this remains an open field of investigation.

The emergence of Ap\'ery-like numbers within the mathematical structure of these walks is another intriguing but
obscure issue. At the present level of our understanding this is something of a mystery, or curiosity. It would be satisfying
to have a better understanding of the relation between random walks and Ap\'ery numbers, with an eye to possible applications
in the mathematics of $\zeta$-functions and/or statistical models.

Finally, the Hamiltonians $H_m$ and ${\tilde H}_m$ presented above are all Hermitian and thus have a real spectrum,
generalizing the corresponding spectrum of the Hofstadter model that leads to the celebrated ``butterfly'' fractal
structure. It is expected that the spectrum of all the above models will have a similarly fractal structure. The shape
and eigenvalue statistics of the spectrum of these generalized models is an intriguing topic for further research.

\vskip 0.25cm

\noindent
{\bf{Acknowledgments}}

\noindent
S.O. acknowledges interesting discussions with Olivier Giraud, in particular regarding (\ref{sonice}) and (\ref{top}). He also thanks Stephan Wagner for mentioning the relation of the  triangular lattice chiral walks of subsection (\ref{krew}) to Kreweras walks. 
A.P. acknowledges the hospitality of LPTMS, CNRS at Universit\'e Paris-Saclay (Facult\'e des Sciences d'Orsay),
where this work was initiated.
A.P.'s research was partially supported by
NSF under grant 1519449 and by an ``Aide Investissements d'Avenir'' LabEx PALM grant (ANR-10-LABX-0039-PALM).

  \section{Appendix }
 {\subsection{\hskip -0.3cm Triangular $r=4$: {${{{{b}}}_{p/q}}(k)\hskip -0.1cm=\hskip -0.1cm\big(2\sin(\pi k p/q)\big)\big(2\sin(\pi (k+1) p/q)\big)\big(2\sin(\pi (k+2) p/q)\big)\big(2\sin(\pi (k+3) p/q)\big)$ \label{tri4}}}}
   Likewise
  \begin{align} &{1\over q}\sum _{k=1}^q {{{{b}}}_{p/q}}^{{l_1}}(k){{{{b}}}_{p/q}}^{l_2}(k+1)\ldots {{{{b}}}_{p/q}}^{l_{j}}(k+j-1)=\sum_{A=-\infty\atop A\;{\rm even}}^{\infty}\rme^{\rmi\pi A p/q}\nonumber\\&\sum_{k_3=-(l_1+l_2+l_3)/2}^{(l_1+l_2+l_3)/2}\sum_{k_4=-(l_1+l_2+l_3+l_4)/2}^{(l_1+l_2+l_3+l_4)/2} 
\ldots \sum_{k_j = -(l_{j-3}+l_{j-2}+l_{j-1}+l_j)/2}^{(l_{j-3}+l_{j-2}+ l_{j-1}+l_j)/2}\sum_{k_{j+1}= -(l_{j-2}+l_{j-1}+l_j)/2}^{(l_{j-2}+l_{j-1}+l_{j})/2}\sum_{k_{j+2}= -(l_{j-1}+l_j)/2}^{(l_{j-1}+l_{j})/2}\sum_{k_{j+3}= -l_j/2}^{l_{j}/2}\nonumber\\&{l_1\choose l_1/2 +A/2+\sum_{i=3}^{j+3}(i-2)k_i}{l_1+l_2\choose (l_1+l_2)/2 -A/2-\sum_{i=3}^{j+3}(i-1)k_i}{l_1+l_2+l_3\choose (l_1+l_2+l_3)/2 +k_3}\nonumber\\&\prod_{i=4}^{j}{l_{i-3}+l_{i-2}+l_{i-1}+l_i\choose (l_{i-3}+l_{i-2}+l_{i-1}+l_i)/2+k_i}{l_{j-2}+l_{j-1}+l_j\choose (l_{j-2}+l_{j-1}+l_j)/2+k_{j+1}}{l_{j-1}+l_j\choose (l_{j-1}+l_j)/2+k_{j+2}}{l_j\choose l_j/2+k_{j+3}}\label{triangular4}\end{align} 
with overall counting 
\[{4(l_1+l_2+\ldots+l_j)\choose 2(l_1+l_2+\ldots+l_j)}\]
One notes that as in previous cases the binomial multiple  sum (\ref{triangular4})  is nothing but  the trigonometric  integral
\begin{align} &{1\over q}\sum _{k=1}^q {{{{b}}}_{p/q}}^{{l_1}}(k){{{{b}}}_{p/q}}^{l_2}(k+1)\ldots {{{{b}}}_{p/q}}^{l_{j}}(k+j-1)= \nonumber\\&\int_0^{1}
 {dt}\bigg(2\sin\big(\pi t\big)\bigg)^{ l_1}\bigg(2\sin\big(\pi t+\pi p/q\big)\bigg)^{l_{1}+ l_2}\bigg(2\sin\big(\pi t+\pi 2 p/q\big)\bigg)^{l_{1}+ l_2+l_3}\nonumber\\&\prod_{i=4}^j 
 \bigg(2\sin\big(\pi t+\pi(i-1)p/q\big)\bigg)^{l_{i-3}+l_{i-2}+l_{i-1}+ l_i}\bigg(2\sin\big(\pi t+\pi j p/q\big)\bigg)^{ l_{j-2}+l_{j-1}+l_j}\nonumber\\&\bigg(2\sin\big(\pi t+\pi (j+1) p/q\big)\bigg)^{l_{j-1}+l_j}\bigg(2\sin\big(\pi t+\pi (j+2) p/q\big)\bigg)^{l_j}\nonumber\end{align}
 under the provision that $2(l_1+\ldots+l_j)<q$.
 
	% {\color{red} {\bf question : why the bounds on $A$ as soon as  $g\ge 4$  are $r/2(=2)$ times the bounds for $r=2$ (see (\ref{triangular})) but they do not seem to be so for $g=2$ and $g=3$?} (Haven't thought about this)}

Clearly for a general $r$ the spectral function
${{{{b}}}_{p/q}}(k)=\big(2\sin(\pi k p/q)\big)\big(2\sin(\pi (k+1) p/q)\big)\ldots\big(2\sin(\pi (k+r-1) p/q)\big)$
 can be treated  along the same lines as in  subsections (\ref{tri2}) and (\ref{tri3})  and above.
 
  {\subsection{Another triangular chiral walks generalization:\\ {${{{{b}}}_{p/q}}(k)=\big(2\sin(\pi k p/q)\big)^{r/2}\big(2\sin(\pi (k+1) p/q)\big)^{r/2}$ \label{nono0}}
 with $r$  even   } }
 
 When ${{b}_{p/q}}(k)=\big(2\sin(\pi k p/q)\big)^2$  we have seen that (\ref{33}), rewritten  as
  \begin{align} &{1\over q}\sum _{k=1}^q {{{{b}}}_{p/q}}^{{l_1}}(k){{{{b}}}_{p/q}}^{l_2}(k+1)\ldots {{{{b}}}_{p/q}}^{l_{j}}(k+j-1)=\sum_{A=-\infty\atop A\;{\rm even}}^{\infty}\rme^{ \rmi \pi A p/q}\nonumber\\&\sum_{k_3= -l_3}^{l_3 
 }\ldots \sum_{k_{j}= -l_j}^{l_{j}}{2l_1\choose l_1 +A/2+\sum_{i=3}^{j}(i-2)k_i}{2l_2\choose l_2 -A/2-\sum_{i=3}^{j}(i-1)k_i}\prod_{i=3}^{j}{2l_i\choose l_i+k_i}\nonumber\end{align} 
 generalizes for ${{b}_{p/q}}(k)=\big(2\sin(\pi k p/q)\big)^r$  and  $r$ is even  to   
\begin{align}&{1\over q}\sum _{k=1}^q {{{{b}}}_{p/q}}^{{l_1}}(k){{{{b}}}_{p/q}}^{l_2}(k+1)\ldots {{{{b}}}_{p/q}}^{l_{j}}(k+j-1)=\sum_{A=-\infty\atop A\;{\rm even}}^{\infty}\rme^{ \rmi \pi A p/q}\nonumber\\&\sum_{k_3= -{rl_3/ 2}}^{{rl_3/2} 
 }\ldots \sum_{k_{j}= -{rl_j/2}}^{{rl_{j}/ 2}}{rl_1\choose {rl_1/ 2} +A/2+\sum_{i=3}^{j}(i-2)k_i}{rl_2\choose {rl_2/2} -A/2-\sum_{i=3}^{j}(i-1)k_i}\prod_{i=3}^{j}{rl_i\choose {rl_i/ 2}+k_i}\nonumber\end{align} 

\noindent Likewise,  when ${{{{b}}}_{p/q}}(k)=\big(2\sin(\pi k p/q)\big)\big(2\sin(\pi (k+1) p/q)\big)$,  equation (\ref{triangular}),
 \begin{align} &{1\over q}\sum _{k=1}^q {{{{b}}}_{p/q}}^{{l_1}}(k){{{{b}}}_{p/q}}^{l_2}(k+1)\ldots {{{{b}}}_{p/q}}^{l_{j}}(k+j-1)\nonumber\\&=\sum_{A=-\infty\atop A\;{\it same}\;{ \it parity}\; l_1+l_2+\ldots+l_j}^{\infty}\rme^{\rmi \pi A p/q}\sum_{k_3=-(l_2+l_3)/2}^{(l_2+l_3)/2} 
\ldots \sum_{k_j = -(l_{j-1}+l_j)/2}^{( l_{j-1}+l_j)/2}\sum_{k_{j+1}= -l_j/2}^{l_{j}/2}\nonumber\\&{l_1\choose {l_1/2 +A/2+\sum_{i=3}^{j+1}(i-2)k_i}}{l_1+l_2\choose (l_1+l_2)/2 -A/2-\sum_{i=3}^{j+1}(i-1)k_i}\nonumber\\&\prod_{i=3}^{j}{l_{i-1}+l_i\choose (l_{i-1}+l_i)/2+k_i}{l_j\choose l_j/2+k_{j+1}}\nonumber\end{align}
  generalizes for ${{{{b}}}_{p/q}}(k)=(\big(2\sin(\pi k p/q)\big)^{r/2}\big(2\sin(\pi (k+1) p/q)\big)^{r/2}$ 
 and $r$  even to
 
\begin{align} &{1\over q}\sum _{k=1}^q {{{{b}}}_{p/q}}^{{l_1}}(k){{{{b}}}_{p/q}}^{l_2}(k+1)\ldots {{{{b}}}_{p/q}}^{l_{j}}(k+j-1)\nonumber\\&=\sum_{A=-\infty\atop A\;{\it same}\;{ \it parity}\; r(l_1+l_2+\ldots+l_j)/2}^{\infty}\rme^{ \rmi \pi A p/q}\sum_{k_3=-r(l_2+l_3)/4}^{r(l_2+l_3)/4} 
\ldots \sum_{k_j = -r(l_{j-1}+l_j)/4}^{r( l_{j-1}+l_j)/4}\sum_{k_{j+1}= -rl_j/4}^{rl_{j}/4}\nonumber\\&{rl_1/2\choose {rl_1/4 +A/2+\sum_{i=3}^{j+1}(i-2)k_i}}{r(l_1+l_2)/2\choose r(l_1+l_2)/4 -A/2-\sum_{i=3}^{j+1}(i-1)k_i}\nonumber\\&\prod_{i=3}^{j}{r(l_{i-1}+l_i)/2\choose r(l_{i-1}+l_i)/4+k_i}{rl_j/2\choose rl_j/4+k_{j+1}}\nonumber\end{align}

\vspace{2cm}

\subsection{Regarding (\ref{sonice}): summing over $A$ odd  when $r$ is even \label{nono}}
So far one has considered the $r(l_1 +l_2 +\ldots+l_j)$ even cases so that the $q\to\infty$  limit in the trigonometric sum ({\ref{11}) yields an overall binomial counting  which is an integer
and contributes as such to the overall counting of closed lattice walks. We have seen that
this trigonometric sum can be rewritten as a multiple binomial sum of the type (\ref{sonice}) or (\ref{triangular}) with some constraints on the evenness or oddness of the $A$'s (and additionnally of $ l_1 +l_2 +\ldots+l_j $ in the case r odd). In the $r(l_1 +l_2 +\ldots+l_j)$ odd cases, on the other hand, (\ref{11}) would not rewrite anymore as a multiple binomial sum. 

Still, and quite generally, one could take the binomial multiple sums (\ref{sonice}) (and likewise (\ref{triangular})) at face value for all possible entries $A$ even or odd and $l_1 + l_2 + \ldots+l_j$  even or odd.
In the $r$ even case we already know that  the $A$ even summation in (\ref{sonice})  has a finite range and yields  exactly  the overall integer counting binomial. The $A$ odd summation happens to yield again the same overall binomial but with each term in the sum a rational number times $1/ \pi^2$  and  an infinite summation range. The $1/\pi^2$ factor  comes from the first two binomials in (\ref{sonice}) due the relaxation of the constraint that their entries  be integers  (since $A$ is now odd).  Likewise in the $r$ odd case, when $l_1+l_2+\ldots+l_j$ is even, we already know that $A$ even or odd summations, depending on the parity of  $l_1+l_3+\ldots$, have a  finite range and yield  the usual overall integer counting binomial; it is still true that  summing over  $A$  even with $l_1+l_3+\ldots$ odd or on $A$ odd with $l_1+l_3+\ldots$  even would yield the same overall counting binomial with again terms $1/\pi^2$ times rational numbers and  an infinite summation range. Finally when both $r$ and $l_1+l_2+\ldots+l_j$ are odd,  $A$ even and odd summations have finite range to   yield   the overall  binomial which is in this case $1/\pi$ times a rational number.  In all these instances   the coefficients  sum up to  ${r(l_1+l_2+\ldots+l_j)\choose r(l_1+l_2+\ldots+l_j)/2}$  for both   $A$ even or odd summations, with finite or infinite ranges depending on the situation. 

\noindent To better understand these weird $A$-summations, let us first focus on the regular $A$-summations 
and consider the LHS of (\ref{top}) i.e., the binomial multiple sum 
\begin{align} \sum_{k_3= -{rl_3/ 2}}^{{rl_3/2} 
 }\hskip -0.1cm\cdots \hskip -0.1cm\sum_{k_{j}= -{rl_j/2}}^{{rl_{j}/ 2}}{rl_1\choose {rl_1/ 2} +A/2+\sum_{i=3}^{j}(i-2)k_i}{rl_2\choose {rl_2/2} -A/2-\sum_{i=3}^{j}(i-1)k_i}\prod_{i=3}^{j}{rl_i\choose {rl_i/ 2}+k_i}\nonumber \end{align}
One wishes to  go backward and get the double integral in the  RHS of (\ref{top}), which, when   
 %\[{1\over 2}\int_0^{2}{dt'}\int_0^{1}{dt}\prod_{i=1}^j \bigg(2\sin\big(\pi t+\pi(i-1)t'\big)\bigg)^{r l_i}e^{i\pi At'}\]
  summed   over $A$,  directly yield the overall counting binomial
 \[{r(l_1+l_2+\ldots+l_j)\choose r(l_1+l_2+\ldots+l_j)/2}\]

\noindent For simplicity  let us  consider the case $r$ even: since $r$ is even, all the $k_i$'s $i=3,\ldots,j $ are integers, and since we know that $A$ has then to be even (see below  (\ref{sonice})), in the first two binomials  both $rl_{1/ 2} +A/2+\sum_{i=3}^{j}(i-2)k_i$ and $ {rl_2/2} -A/2-\sum_{i=3}^{j}(i-1)k_i$  are integers.
Using that for an integer  $n$
\[\int_0^1 dt e^{2 i \pi (k-n) t}\] is the Kronecker $\delta(k,n)$ meaning  
\[\sum_{k=-\infty\atop k\;{\it  integer}}^{\infty}\delta(k,n)f(k)=f(n) \]
we can rewrite these binomials  as
\be {rl_1\choose {rl_1/ 2} +A/2+\sum_{i=3}^{j}(i-2)k_i}=\sum_{k_1=-rl_1/2\atop k_1\;{\it  integer}}^{rl_1/2}\int_0^1 dt \rme^{2 \rmi \pi \big(k_1-(A/2+\sum_{i=3}^{j}(i-2)k_i)\big) t} {rl_1\choose {rl_1/ 2} +k_1} \nonumber\ee

\be {rl_2\choose {rl_2/ 2} -A/2-\sum_{i=3}^{j}(i-1)k_i}=\sum_{k_2=-rl_2/2\atop k_2\;{\it  integer}}^{rl_2/2}\int_0^1 dt' \rme^{2 \rmi \pi \big(k_2+A/2+\sum_{i=3}^{j}(i-1)k_i\big) t'} {rl_2\choose {rl_2/ 2} +k_2} \nonumber\ee
where the summations are restricted to $[-rl_1/2,rl_1/2]$ and $[-rl_2/2,rl_2/2]$  since there is no point to sum outside these intervals  where the binomials trivially vanish.
 So the LHS  of (\ref{top}) becomes
\begin{align} \sum_{k_1= -{rl_1/ 2}\atop k_1\;{\it  integer}}^{{rl_1/2} 
 }\hskip -0.1cm\cdots \hskip -0.1cm\sum_{k_{j}= -rl_j/2\atop k_j\;{\it  integer}}^{{rl_{j}/ 2}}\int_0^1 dt \rme^{2 \rmi \pi \big(k_1-(A/2+\sum_{i=3}^{j}(i-2)k_i)\big) t}\int_0^1 dt' \rme^{2 \rmi \pi \big(k_2+A/2+\sum_{i=3}^{j}(i-1)k_i\big) t'}\prod_{i=1}^{j}{rl_i\choose {rl_i/ 2}+k_i}\nonumber \end{align}
  which is
  \begin{align} \sum_{k_1= -{rl_1/ 2}\atop k_1\;{\it  integer}}^{{rl_1/2} 
 }\hskip -0.1cm\cdots \hskip -0.1cm\sum_{k_{j}= -{rl_j/2}\atop k_j\;{\it  integer}}^{{rl_{j}/ 2}}\int_0^1 dt\int_0^1 dt' \rme^{ \rmi \pi A(t'-t)}\prod_{i=1}^{j}{rl_i\choose {rl_i/ 2}+k_i} \rme^{2 \rmi \pi k_i\big((i-1)t'-(i-2)t)\big) }\nonumber \end{align}
 i.e., since obviously
 \[\sum_{k_{i}= -{rl_i/2}\atop k_i\;{\it  integer}}^{{rl_{i}/ 2}}{rl_i\choose {rl_i/ 2}+k_i}\rme^{2 \rmi \pi k_i\big((i-1)t'-(i-2)t\big)}=\bigg(2\cos\big(\pi\left((i-1)t'-(i-2)t\right)\big)\bigg)^{rl_i}\]
  and calling $t'-t=t"$, we obtain\footnote{Or equivalently  as in the RHS of (\ref{top})
  \begin{align}{1\over 2} \int_0^1 dt\int_0^2 dt" \rme^{ \rmi \pi At"}\prod_{i=1}^{j}\bigg(2\sin\big(\pi\left((i-1)t"+t\right)\big)\bigg)^{rl_i}\nonumber \end{align}}
 \begin{align} \int_0^1 dt\int_0^1 dt" \rme^{ \rmi \pi At"}\prod_{i=1}^{j}\bigg(2\cos\big(\pi\left((i-1)t"+t\right)\big)\bigg)^{rl_i}\nonumber \end{align}
 We have   to sum over $A$  even: since  $\sum_{A\;{\it even}}\rme^{ \rmi \pi At"}=\sum_{n=-\infty} ^{\infty}\delta(t",n) $  
   \begin{align}\sum_{A\;{\it even}} \int_0^1 dt\int_0^1 dt" \rme^{ \rmi \pi At"}\prod_{i=1}^{j}\bigg(2\cos\big(\pi\left((i-1)t"+t\right)\big)\bigg)^{rl_j}&=\int_0^1 dt \big(2\cos(\pi t)\big)^{r(l_1+l_2+\ldots+l_j)}\nonumber\\&={r(l_1+l_2+\ldots+l_j)\choose r(l_1+l_2+\ldots+l_j)/2}\label{ok} \end{align}
   where the overall binomial counting has been obtained as expected.
  
\noindent Now still assuming $r$ being even, so that all the $k_i$'s $i=3,\ldots,j $ are integers, let us insist that the summation over  $A$  be on $A$ odd so that both ${rl_1/ 2} +A/2+\sum_{i=3}^{j}(i-2)k_i$ and $ {rl_2/2} -A/2-\sum_{i=3}^{j}(i-1)k_i$  are half-integers.
 Using that for an half-integer  $n/2$
\[\int_0^1 dt e^{2 i \pi (k-n/2) t}\] is the Kronecker $\delta(k,n/2)$ meaning 
\[\sum_{k=-\infty\atop k\;{\it half}\;{\it  integer}}^{\infty}\delta(k,n/2)f(k)=f(n/2) \]
 we rewrite the same two  binomials  as
\be {rl_1\choose {rl_1/ 2} +A/2+\sum_{i=3}^{j}(i-2)k_i}=\sum_{k_1=-\infty\atop k_1\;{\it half}\;{\it  integer}}^{\infty}\int_0^1 dt \rme^{2 \rmi \pi \big(k_1-(A/2+\sum_{i=3}^{j}(i-2)k_i)\big) t} {rl_1\choose {rl_1/ 2} +k_1} \nonumber\ee
\be {rl_2\choose {rl_2/ 2} -A/2-\sum_{i=3}^{j}(i-1)k_i}=\sum_{k_2=-\infty \atop k_2\;{\it half}\;{\it  integer}}^{\infty}\int_0^1 dt' \rme^{2 \rmi \pi \big(k_2+A/2+\sum_{i=3}^{j}(i-1)k_i\big) t'} {rl_2\choose {rl_2/ 2} +k_2} \nonumber\ee
 Doing the same manipulations as above except for the first two binomials   the LHS of (\ref{top}) then becomes 
\begin{align} &\int_0^1 dt\int_0^1 dt"\sum_{k_1=-\infty\atop k_1\;{\it half}\;{\it  integer}}^{\infty}\sum_{k_2=-\infty\atop k_2\;{\it half}\;{\it  integer}}^{\infty}{rl_1\choose {rl_1/ 2}+k_1}{rl_2\choose {rl_2/ 2}+k_2}\nonumber\\& \rme^{ \rmi \pi At"}\rme^{ 2\rmi \pi (k_1+k_2)t}\rme^{ 2\rmi \pi k_2t"}\prod_{i=3}^{j}\bigg(2\cos\big(\pi\left((i-1)t"+t\right)\big)\bigg)^{rl_i}\nonumber\end{align}
 Summing over all $A$ odd i.e., over $A+2k_2$ even --since $k_2$ is an half integer--  yields again a Kronecker enforcing   $t"=0$ so that after summation one obtains
 \begin{align} &\int_0^1 dt\sum_{k_1=-\infty\atop k_1\;{\it half}\;{\it  integer}}^{\infty}\sum_{k_2=-\infty\atop k_2\;{\it half}\;{\it  integer}}^{\infty}{rl_1\choose {rl_1/ 2}+k_1}{rl_2\choose {rl_2/ 2}+k_2}\rme^{ 2\rmi \pi (k_1+k_2)t}\big(2\cos(\pi t)\big)^{r(l_3+\ldots+l_j)}\nonumber\end{align}
 Comparing with (\ref{ok}) we see that in order to get the same overall  binomial  counting
 everything boils down to showing that in the same way that obviously \begin{align}&\sum_{k_1=-rl_1/2\atop k_1\;{\it  integer}}^{rl_1/2}\sum_{k_2=-rl_2/2\atop k_2\;{\it  integer}}^{rl_2/2}{rl_1\choose {rl_1/ 2}+k_1}{rl_2\choose {rl_2/ 2}+k_2}\rme^{ 2\rmi \pi (k_1+k_2)t}\nonumber\\&= \big(2\cos(\pi t)\big)^{r(l_1+l_2)}\label{oula}\end{align} holds,
 \begin{align} &\sum_{k_1=-\infty\atop k_1\;{\it half}\;{\it  integer}}^{\infty}\sum_{k_2=-\infty\atop k_2\;{\it half}\;{\it  integer}}^{\infty}{rl_1\choose {rl_1/ 2}+k_1}{rl_2\choose {rl_2/ 2}+k_2}\rme^{ 2\rmi \pi (k_1+k_2)t}\nonumber\\&= \big(2\cos(\pi t)\big)^{r(l_1+l_2)}\nonumber\end{align} should  also hold.

\noindent  To show this let us focus on the trivial  identity (\ref{oula}) 
which is nothing but 
\begin{align}&\sum_{k_1=-rl_1/2\atop k_1\;{\it  integer}}^{rl_1/2}\sum_{k_2=-rl_2/2\atop k_2\;{\it  integer}}^{rl_2/2}{rl_1\choose {rl_1/ 2}+k_1}{rl_2\choose {rl_2/ 2}+k_2}\rme^{ 2\rmi \pi (k_1+k_2)t}=\sum_{k=-r(l_1+l_2)/2\atop k\;{\it  integer}}^{r(l_1+l_2)/2}{r(l_1+l_2)\choose {r(l_1+l_2)/ 2}+k}\rme^{ 2\rmi \pi k t}\nonumber\end{align}  
or equivalently,
harmlessly relaxing the range of $k_1, k_2$ and $k$  summations, 
\begin{align}&\sum_{k_1=-\infty\atop k_1\;{\it integer}}^{\infty}\sum_{k_2=-\infty\atop k_2\;{\it integer}}^{\infty}{rl_1\choose {rl_1/ 2}+k_1}{rl_2\choose {rl_2/ 2}+k_2}\rme^{ 2\rmi \pi (k_1+k_2)t}=\sum_{k=-\infty\atop k\;{\it integer}}^{\infty}{r(l_1+l_2)\choose {r(l_1+l_2)/ 2}+k}\rme^{ 2\rmi \pi k t}\label{mybeauty}\end{align}
Let us to rederive it in an other way : defining $k=k_1+ k_2$  we can rewrite
\begin{align}&\sum_{k_1=-\infty\atop k_1\;{\it integer}}^{\infty}\sum_{k_2=-\infty\atop k_2\;{\it integer}}^{\infty}{rl_1\choose {rl_1/ 2}+k_1}{rl_2\choose {rl_2/ 2}+k_2}\rme^{ 2\rmi \pi (k_1+k_2)t}\nonumber\\&= \sum_{k_+=-\infty\atop k\;{\it integer}}^{\infty}\sum_{k_1=-\infty\atop k_1\;{\it integer}}^{\infty}{rl_1\choose {rl_1/ 2}+k_1}{rl_2\choose {rl_2/ 2}+k-k_1}\rme^{ 2\rmi \pi  kt}\nonumber\end{align}
Thanks to the  Chu-Vandermonde identity
\begin{align} &{l_1+l_2\choose l'_1+l'_2}=\sum_{k_1=-max(l'_1,l'_2)\atop k_1\;{\it integer}}^{max(l'_1,l'_2)}{{l_1}\choose {l'_1}+k_1}{ {l_2} \choose {l'_2}-k_1 }\nonumber\end{align}  we conclude that
we indeed  recover (\ref{mybeauty}).

\noindent  It is clear that the same conclusion can be reached  when $k_1$ and $k_2$ are now both half integers namely
\begin{align}&\sum_{k_1=-\infty\atop k_1\;{\it half}\;{\it integer}}^{\infty}\sum_{k_2=-\infty\atop k_2\;{\it half}\;{\it integer}}^{\infty}{rl_1\choose {rl_1/ 2}+k_1}{rl_2\choose {rl_2/ 2}+k_2}\rme^{ 2\rmi \pi (k_1+k_2)t}\nonumber\\&=\sum_{k=-\infty\atop k\;{\it integer}}^{\infty}{r(l_1+l_2)\choose {r(l_1+l_2)/ 2}+k}\rme^{ 2\rmi \pi k t}\label{sosonice}\end{align} Indeed $k_1$ and $k_2$ being both half integers then $k=k_1+ k_2$  is again an  integer  so we can write 
\begin{align}&\sum_{k_1=-\infty\atop k_1\;{\it half}\;{\it integer}}^{\infty}\sum_{k_2=-\infty\atop k_2\;{\it half}\;{\it integer}}^{\infty}{rl_1\choose {rl_1/ 2}+k_1}{rl_2\choose {rl_2/ 2}+k_2}\rme^{ 2\rmi \pi (k_1+k_2)t}\nonumber\\&= \sum_{k=-\infty\atop k\;{\it integer}}^{\infty}\sum_{k_1=-\infty\atop k_1\;{\it half}\;{\it integer}}^{\infty}{rl_1\choose {rl_1/ 2}+k_1}{rl_2\choose {rl_2/ 2}+k-k_1}\rme^{ 2\rmi \pi  kt}\nonumber\end{align}
Thanks to
the  generalized Chu-Vandermonde identity 
\begin{align} &{l_1+l_2\choose l'_1+l'_2}=\sum_{k_1=-\infty\atop k_1\;{\it half}\;{\it integer}}^{\infty}{{l_1}\choose {l'_1}+k_1}{ {l_2} \choose {l'_2}-k_1 }\nonumber\end{align}
we reach  indeed the identity (\ref{sosonice})  for the half integers summations.  
 From which  it directly follows that in the presence of the additional  $\big(2\cos(\pi t)\big)^{r(l_3+\ldots+l_j)}$ term   integrating over $t$ from $0$ to $1$ 
  one ends up getting again the same overall binomial counting, as desired.

  \begin{figure}
%\begin{center}
%\vspace{1cm}
\vskip -5cm\includegraphics[scale=1.2]{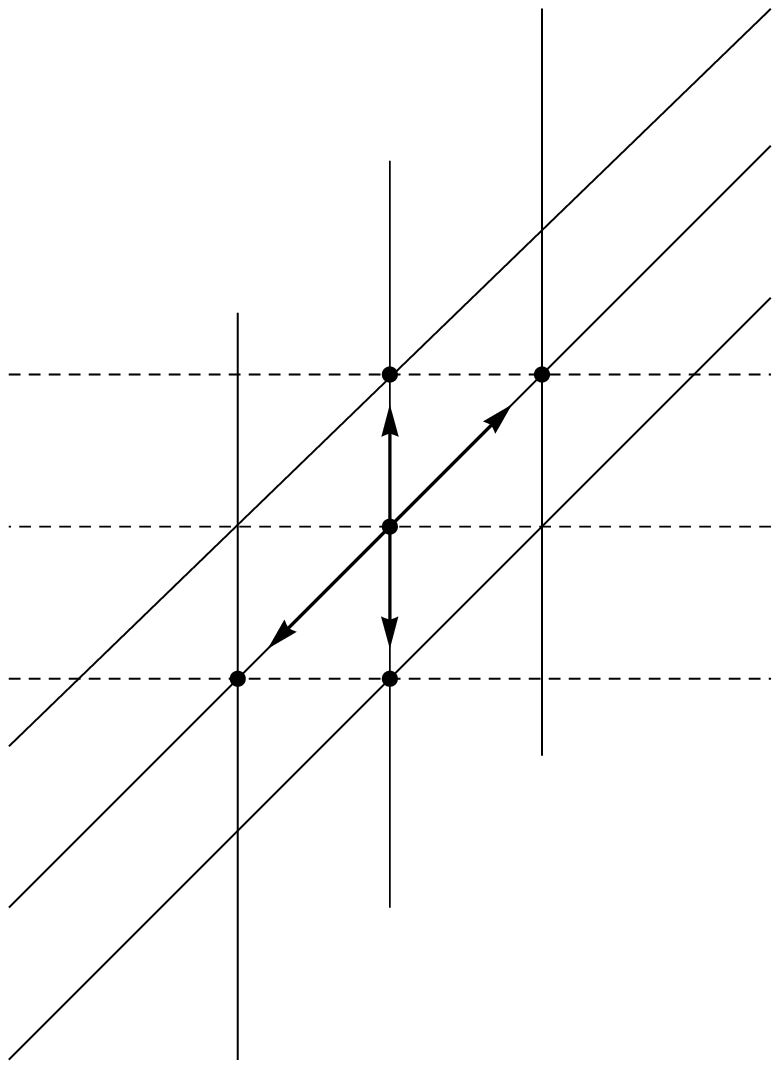}
\vskip -15cm
\caption{The lattice in (\ref{myHof}).} 
%\end{center}
\end{figure}

\begin{figure}
%\begin{center}
\hskip 2.9cm\includegraphics[scale=.5]{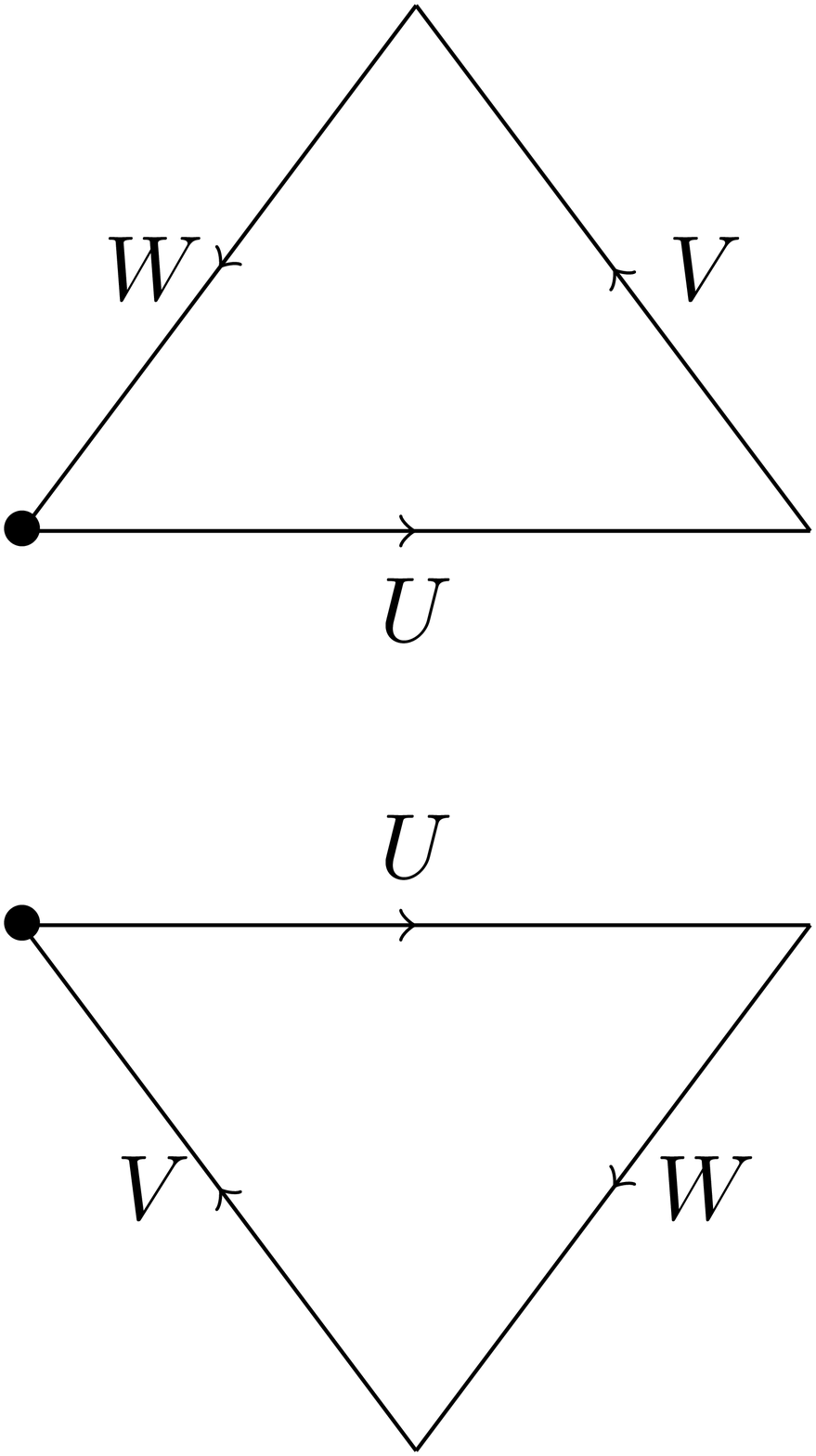}\label{fig1}
\vskip 1cm
\caption{U, V and W are the three possible hoppings on the triangular lattice. As an illustration two chiral walks going around {\it up-vertex} and {\it down-vertex} triangular cells starting from the  black bullet lattice site.}
%\end{center}
\end{figure}

\begin{figure}
%\begin{center}
%\hspace{2cm}
%\vspace{2cm}
\vskip -5cm
\hskip 0.9cm\includegraphics[scale=.7]{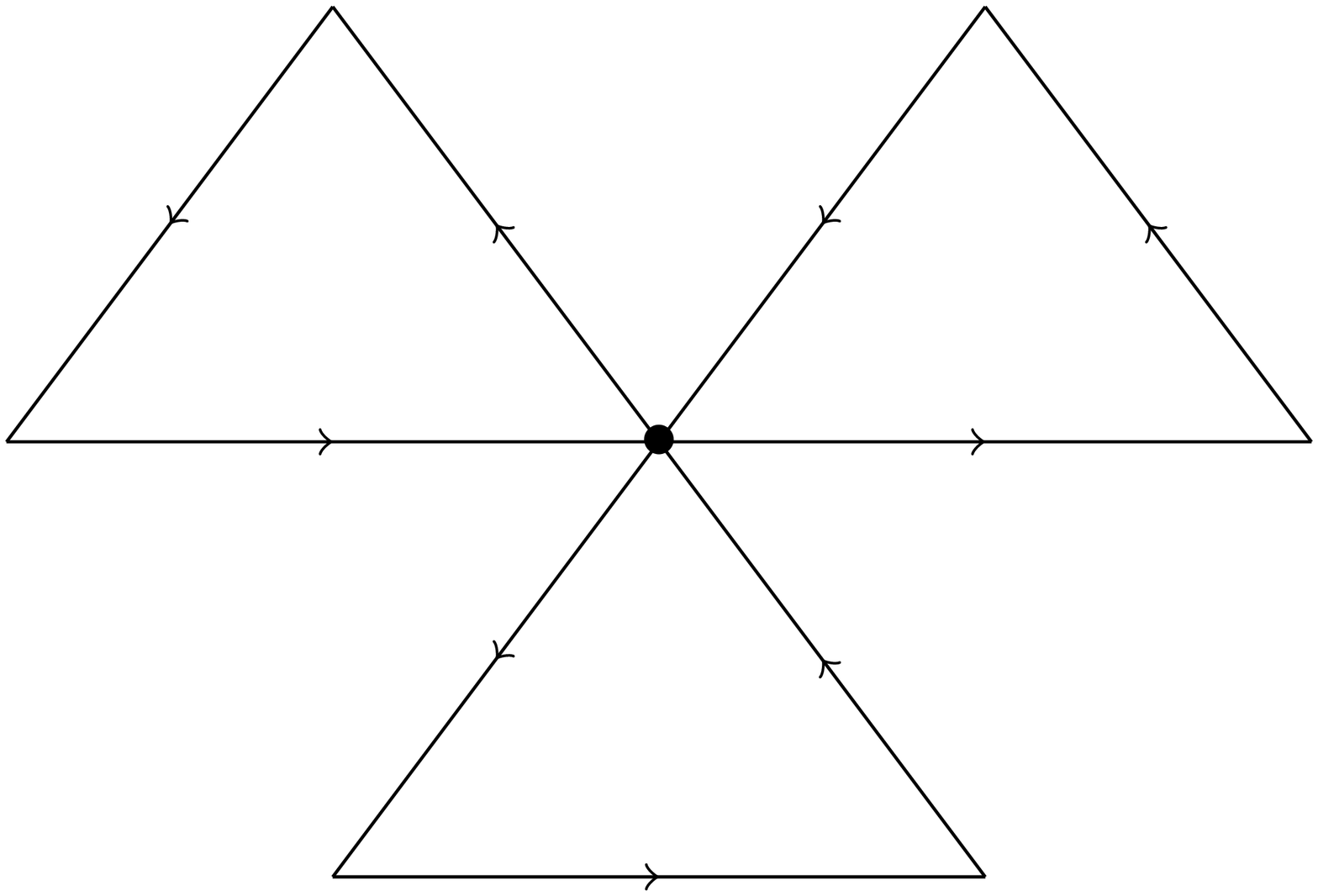}\label{fig2}
\vskip -3cm
\caption{Three of the 6 possible chiral walks starting from the same black bullet lattice site. Only the 3 outgoing arrows
represent possible motions from the original site.}
%\end{center}
\end{figure}

\begin{figure}
%\begin{center}
\vskip -1cm
\hskip 0.9cm\includegraphics[scale=.7]{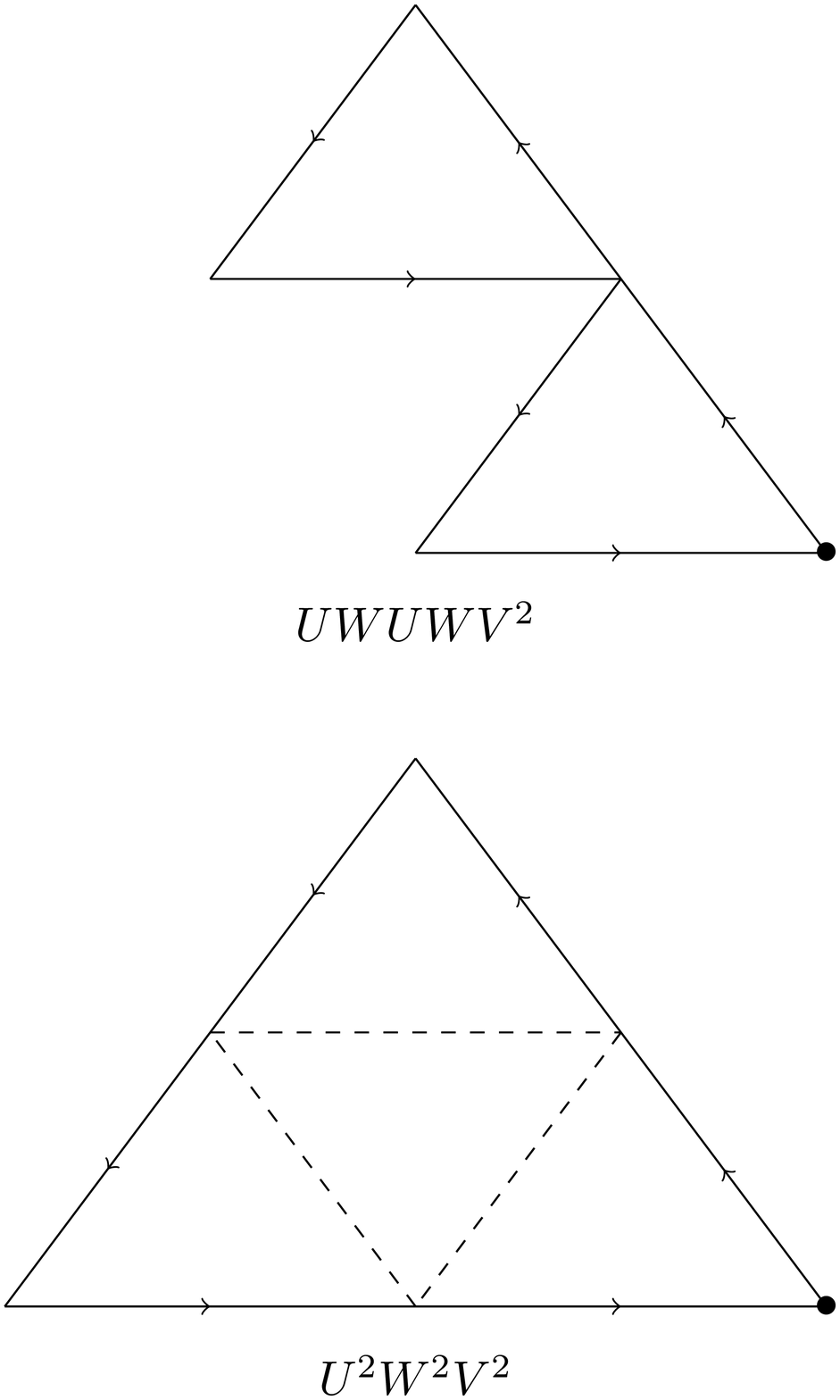}\label{fig3}
\vskip 0.7cm
\caption{$UWUWV^2$  and $U^2 W^2 V^2$ walks.} 
%\end{center}
\end{figure}


\begin{thebibliography}{99}

%\bibitem{old} {see e.g., }

\bibitem{nous}  S. Ouvry and S. Wu, ``The algebraic area of closed lattice random walks", Journal of Physics A: Mathematical and Theoretical, Volume 52, Number 25 (2019).

\bibitem{Hofstadter} D.R. Hofstadter, ``Energy levels and wave functions of Bloch electrons in rational and irrational magnetic fields", Phys. Rev. B {\bf 14} (1976) 2239.

\bibitem{poly} S. Ouvry and A. Polychronakos, ``Exclusion statistics and lattice random walks", NPB[FS] 948 (2019)114731.

%\bibitem{Kreft} C. Kreft, ``Explicit Computation of the Discriminant for the Harper Equation with Rational Flux", %SFB 288 Preprint No. 89 (1993).

\bibitem{Haldane} F.D.M. Haldane, ``Fractional statistics in arbitrary dimensions: A generalization of the Pauli principle", Phys. Rev. Lett. {\bf 67} (1991) 937–940; see also Y.S. Wu, ``Statistical distribution for generalized ideal gas of fractional-statistics particles", Phys. Rev. Lett. 73 (1994) 922–925; A.P. Polychronakos, ``Nonrelativistic bosonization and fractional statistics", Nucl. Phys. {\bf B324} (1989) 597;
 A. Dasni\`eres de Veigy and S. Ouvry, ``Equation of State of an Anyon gas in a Strong Magnetic Field", Phys. Rev. Lett. {\bf 72} (1994) 600.
\bibitem{kreweras}  see e.g., O. Bernardi, ``Bijective counting of Kreweras walks  and loopless triangulations"  Journal of Combinatorial Theory - Series A, Vol 114(5) (2007)  931-956. 




\end{thebibliography}
\end{document}